\documentclass{article}
\usepackage{latexsym}
\usepackage{graphicx}
\usepackage{color}
\usepackage{mathrsfs}
\usepackage{amsmath,amssymb}
\usepackage{amsthm}
\usepackage{mathtools}
\usepackage{breqn}
\usepackage{bm}
\usepackage{bbm}
\usepackage{ascmac}
\usepackage{braket}
\usepackage{geometry}
\usepackage{here}
\theoremstyle{plain}
\usepackage[unicode,bookmarks=true,colorlinks=true]{hyperref}

\numberwithin{equation}{section}
\newtheorem{thm}{Theorem}[section]
\newtheorem{lem}[thm]{Lemma}

\newtheorem{dfn}[thm]{Definition}
\newtheorem{asm}[thm]{Assumption}
\newtheorem{rem}[thm]{Remark}

\mathtoolsset{showonlyrefs}

\usepackage[unicode,bookmarks=true,colorlinks=true]{hyperref}

\providecommand{\keywords}[1]
{	
  \textbf{{Keywords~}} asset pricing model, exponential utility, mean field game, partial observation 
}
\title{\textbf{\Large{Mean field equilibrium asset pricing model under partial observation: An exponential quadratic Gaussian approach}}\thanks{Forthcoming in \textit{Japan Journal of Industrial and Applied Mathematics.} All content presented in this research is solely the opinion of the author and does not represent the views of any institution. The author disclaims any responsibility or liability for any losses or damages resulting from the use of the research.}}
\author{Masashi Sekine\thanks{sekinemasashi@g.ecc.u-tokyo.ac.jp, https://orcid.org/0009-0002-5042-9895, Graduate School of Economics, The University of Tokyo. 7-3-1 Hongo, Bunkyo-ku, Tokyo, 113-0033, JAPAN.}}
\date{April 1, 2025}
\begin{document}

\maketitle
\begin{abstract}
    This paper studies an asset pricing model in a partially observable market with a large number of heterogeneous agents using the mean field game theory. 
    In this model, we assume that investors can only observe stock prices and must infer the risk premium from these observations when determining trading strategies. 
    We characterize the equilibrium risk premium in such a market through a solution to the mean field backward stochastic differential equation (BSDE).
    Specifically, the solution to the mean field BSDE can be expressed semi-analytically by employing an exponential quadratic Gaussian framework. 
    We then construct the risk premium process, which cannot be observed directly by investors, endogenously using the Kalman-Bucy filtering theory.
    In addition, we include a simple numerical simulation to visualize the dynamics of our market model.
\end{abstract}
\keywords

\section{Introduction}
\subsection{Preliminary}
The theory of asset pricing is one of the major interests in financial economics. It examines the formulation of prices in the market at equilibrium, the state where the demand for securities matches the supply. 
Comprehensive overviews of this topic can be found in, for example, Back \cite{backAssetPricingPortfolio2017} and Munk \cite{Munk}. Additionally, we refer to Karatzas \& Shreve \cite{karatzas_methods_1998} [Section 4] for details of the asset pricing in a complete market and Jarrow \cite{jarrow2018continuous} [Part III] for an organized review of the asset pricing in an incomplete market.
\par

Investors generally do not have full access to market information, which necessitates them to infer the risk premium from the observable security price in order to make decisions about their trading strategies. 
This type of problem has intrigued researchers and has led to numerous studies, including the mean-variance hedging (MVH) problem and the utility maximization problem. See, for instance, \cite{pham_meanvariance,MTT_MVH,fujiiMakingMeanvarianceHedging2014} for MVH problem and \cite{lanker_optim_trading, pham_quenez_OP, ManiaSantacroce} for the utility maximization problem.
The key theory behind these partially observable market problems is the stochastic filtering theory. The objective of this theory is to provide the ``best estimate" of the state process based on the observations. 
As a particular case, the linear filtering, developed by Kalman \& Bucy \cite{kalman-bucy}, has been widely used to address these problems. 
Comprehensive literature on the stochastic filtering theory can be found in, for example, Bain \& Crisan \cite{bain_fundamentals_2009} and Liptser \& Shiryayev \cite{LiptserShiryayev}.

\par
Mean field game theory was independently formulated by Lasry \& Lions \cite{lasryMeanFieldGames2007} and Huang, Malham\'{e} \& Caines \cite{huangLargePopulationStochastic2006}, providing a powerful framework for analyzing the problem of multi-agent games. 
Traditional approaches to such problems typically become intractable because of complex interactions among agents, whereas the mean field game theory overcomes this issue by replacing these problems with a stochastic control problem of a single representative agent and a fixed-point problem. 
Carmona \& Delarue \cite{carmonaProbabilisticAnalysisMeanField2013, carmonaForwardBackwardStochastic2015} proposed the probabilistic approach to the mean field problem, involving forward-backward stochastic differential equations (FBSDEs) of McKean-Vlasov type.
The solution of the mean field game is known to yield an $\varepsilon$-Nash equilibrium of the original multi-agent game. Their theory is extensively covered in the two-volume monographs Carmona \& Delarue \cite{carmonaProbabilisticTheoryMean2018,carmonaProbabilisticTheoryMean2018a} with thorough details and applications.

\par
For research on the mean field game theory under partial observation, we refer to Huang, Caines \& Malham\'{e} \cite{huangPO} for an early study of mean field linear quadratic Gaussian (LQG) games with partial information, in which each agent has a local noisy measurement of its own state. Huang, Wang \& Wu \cite{HWW_LQGPO_2016} originally developed a backward mean field LQG game under partial information.
Bensoussan, Feng \& Huang \cite{Bensoussan_et_al_2021} offers an extension for mean field LQG games under partial observation with common noise. 
Huang \& Wang \cite{huangdynamic_optim_2016} investigates dynamic optimization problems of a large-population system and \c{S}en \& Caines \cite{SenCaines_2019} studies a partially observed mean field game with nonlinear cost functionals and dynamics. 
Recent contributions include Li, Nie \& Wu \cite{Li_et_al2023} for a stochastic large-population problem with partial information, where the diffusion term depends also on the control variable, and Li, Li \& Wu \cite{Li_et_al_2024} for problems where agents are coupled through the control average term.

\par
In recent years, there has been an increasing number of studies on asset pricing in financial markets employing the mean field game theory. Evangelista, Saporito \& Thamsten \cite{evangelista2022price} developed an asset pricing model considering liquidity issues using the mean field game theory. 
Fujii \& Takahashi \cite{fujiiMeanFieldGame2022, Fujii-Takahashi_strong, fujii2022equilibrium} presented a mean field price formation model under stochastic order flow.  
Fujii \cite{Fujii-equilibrium-pricing} developed a price formation model that considers market participants of two groups: cooperative and non-cooperative ones.
Fujii \& Sekine \cite{fujiiMeanFieldEquilibriumPrice2023a} introduced a mean field equilibrium pricing model in an incomplete market participated by heterogeneous agents with exponential utility. This model shows that the equilibrium risk premium process is characterized by a novel form of the mean field BSDE and proves its well-posedness under certain conditions using the method of Tevzadze \cite{tevzadzeSolvabilityBackwardStochastic2008}. 
The same authors extended this work in \cite{fujiiMeanFieldEquilibriumAsset2024} by taking the agents' consumption behavior and habit formation into account.
It also considers a mean field BSDE of a similar type and proves its well-posedness. It then introduces an exponential quadratic Gaussian (EQG) approach, in which the aforementioned mean field BSDE admits a semi-analytical solution.
\par

The main contribution of this paper is an extension of \cite{fujiiMeanFieldEquilibriumPrice2023a,fujiiMeanFieldEquilibriumAsset2024} to the case of partial observation under the exponential quadratic Gaussian framework. As mentioned above, we assume that investors can only observe the security price but cannot distinguish between the risk premium process and the Brownian noise. 
Our objective is to derive the market risk premium processes, which cannot be directly observed by agents, endogenously from the optimal behavior of agents and the market clearing condition by using the linear filtering theory. 
As in the previous work, we assume that agents are characterized by exponential-type preferences and adopt self-financing strategies. In addition, we allow agents to have heterogeneity in initial wealth and terminal liability, in contrast to the traditional asset pricing theory which considers a single representative agent.
We employ an exponential quadratic Gaussian formulation, which not only provides a semi-explicit solution of the mean field equilibrium but also allows us to conduct numerical simulations. 
\par

This paper is organized as follows. 
In the rest of Section 1, we introduce the notations for frequently used sets. 
Section 2 presents a formulation of the partially observable market and the utility maximization problem of an agent, along with the derivation of the conditions for an optimal strategy. 
In Section 3, we introduce the asymptotic market clearing condition and consider the relevant mean field BSDE. By associating the BSDE with a system of ordinary differential equations (ODEs), we show that the solution of the BSDE allows a semi-explicit solution. Furthermore, we verify that this solution does indeed characterize the market clearing condition in the large population limit. We then construct the risk premium process under the Kalman-Bucy framework.
Section 4 provides a numerical simulation to visualize this model. The paper concludes in Section 5 with a suggestion for possible extensions.

\subsection{Notations}
In this paper, we shall work on a finite time interval $[0,T]$ for some $T>0$. For a given filtered probability space with usual conditions $(\Omega,\mathcal{F},\mathbb{P},\mathbb{F}~(:=(\mathcal{F}_t)_{t\in[0,T]}))$, a sub $\sigma$-algebra $\mathcal{G}\subset\mathcal{F}$ and a vector space $E$ over $\mathbb{R}$, we use the following notations to describe frequently used sets and function spaces.\\

\noindent
(1) $\mathcal{T}(\mathbb{F})$ is a set of all $\mathbb{F}$-stopping times with values in $[0,T]$.\\

\noindent
(2) $\mathbb{L}^0(\mathcal{G},E)$ is a set of $E$-valued $\mathcal{G}$-measurable random variables. \\

\noindent
(3) $\mathbb{L}^2(\mathbb{P},\mathcal{G},E)$ is a set of $E$-valued $\mathcal{G}$-measurable random variables $\xi$ satisfying $\|\xi\|_2:=\mathbb{E}^{\mathbb{P}}[|\xi|^2]^{\frac{1}{2}}<\infty$.\\

\noindent
(4) $\mathbb{L}^0(\mathbb{F},E)$ is a set of $E$-valued $\mathbb{F}$-progressively measurable stochastic processes. \\

\noindent
(5) $\mathbb{H}^2(\mathbb{P},\mathbb{F},E)$ is a set of $E$-valued $\mathbb{F}$-progressively measurable stochastic processes $X$ satisfying
\[
\|X\|_{\mathbb{H}^2}:=\mathbb{E}^{\mathbb{P}}\Bigl[\int_0^T |X_t|^2 dt\Bigr]^{\frac{1}{2}}<\infty.
\]

\noindent
(6) $\mathbb{S}^2(\mathbb{P},\mathbb{F},E)$ is a set of $E$-valued continuous $\mathbb{F}$-adapted stochastic processes $X$ satisfying
\[
\|X\|_{\mathbb{S}^2}:=\mathbb{E}^{\mathbb{P}}\Bigl[\sup_{t\in[0,T]}|X_t|^2\Bigr]^{\frac{1}{2}}<\infty.
\]

\noindent
(7) $\mathcal{C}([0,T],E)$ is a set of continuous functions $f:[0,T]\to E$. \\

\noindent
(8) $\mathcal{C}^1([0,T],E)$ is a set of once continuously differentiable functions $f:[0,T]\to E$. We simply say ``$f$ is of class $\mathcal{C}^1$'' if relevant sets are obvious. \\

\noindent
(9) We set $\mathbb{R}^n_+:=\{x\in\mathbb{R}^n; x\geq 0\}$ and $\mathbb{R}^n_{++}:=\{x\in\mathbb{R}^n; x> 0\}$ for $n\in\mathbb{N}$, where $x\geq 0$ and $x> 0$ means that all elements of $x$ are nonnegative and strictly positive, respectively. Also, $\mathbb{M}_n$ is a set of real symmetric matrices of size $n\times n$.\par
We may omit the arguments such as $\omega\in\Omega$ and $(\mathbb{P},\mathcal{F},\mathbb{F},E)$ if obvious.

\section{The market with partial observation} 
This section studies an optimal investment problem for a single agent in a partially observable market. It basically follows Fujii \& Sekine \cite{fujiiMeanFieldEquilibriumPrice2023a} by adopting the technique of Hu, Imkeller \& M\"{u}ller \cite{huUtilityMaximizationIncomplete2005a}. 
To deal with the partial observation, we shall mention some results of the filtering theory for completeness.\par

We denote by $(\Omega^0,\mathcal{F}^0,\mathbb{P}^0)$ a complete probability space with a complete and right-continuous filtration $\mathbb{F}^0:=(\mathcal{F}^0_t)_{t\in[0,T]}$ generated by a $d_0$-dimensional standard Brownian motion $W^0:=(W^0_t)_{t\in[0,T]}$, a $k$-dimensional standard Brownian motion $B^0:=(B^0_t)_{t\in[0,T]}$ and an $\mathbb{R}^{d_0}$-valued random variable $\mu_0$. Here, we assume that $W^0$ and $B^0$ are independent. $\mathcal{F}^0_0$ is the completion of $\sigma(\mu_0)$. We set $\mathcal{F}^0 := \mathcal{F}^0_T$. $(\Omega^0,\mathcal{F}^0,\mathbb{P}^0)$ is used to describe the randomness of the financial market. \par
\subsection{Market setup}
The market dynamics and its properties are given in the following assumption.
\begin{asm}~\\
    \label{asm1}
    \textup{(i)} The risk-free interest rate is zero.\\
    \textup{(ii)} There are $d_0$ non-dividend paying risky stocks with price dynamics
    \begin{equation}
        \begin{split}
            \label{stock price-EQG}
            S_t&= S_0 + \int_0^t \mathrm{diag}(S_r)(\mu_rdr + \sigma_r dW^0_r),~~t\in[0,T],
        \end{split}
    \end{equation}
    for $S_0\in\mathbb{R}^{d_0}_{++}$, $\mu := (\mu_t)_{t\in[0,T]}\in\mathbb{H}^2(\mathbb{P}^{0},\mathbb{F}^0,\mathbb{R}^{d_0})$ with $\mu_0\in\mathbb{L}^2(\mathbb{P}^0,\mathcal{F}^0_0,\mathbb{R}^{d_0})$ and a measurable function $\sigma:[0,T]\to \mathbb{R}^{d_0\times d_0}$. \\
    \textup{(iii)} $\sigma_t$ is invertible for all $t\in[0,T]$ and satisfies
    \[
        \underline{\lambda}I_{d_0}\leq (\sigma_t\sigma_t^\top)\leq\overline{\lambda}I_{d_0},~~~~dt\otimes \mathbb{P}^0\text{-}\mathrm{a.e.}
    \]
    for some positive constants $0<\underline{\lambda}\leq\overline{\lambda}$ and $I_{d_0}$, an identity matrix of size $d_0$.\\
    \textup{(iv)} The risk premium process $\theta\in\mathbb{H}^2(\mathbb{P}^{0},\mathbb{F}^0,\mathbb{R}^{d_0})$, defined by $\theta_t = \sigma_t^{-1}\mu_t$ for $t\in[0,T]$, is a process such that the Dol\'{e}ans-Dade exponential $\displaystyle\Bigl\{\mathcal{E}\Bigl(-\int_0^\cdot \theta_s^\top dW^0_s\Bigr)_t; t\in[0,T]\Bigr\}$ is a martingale.
\end{asm}

\begin{rem}~\\
    \label{well-posedness_S}
    Although $\mu$ is unbounded, the well-posedness of the stock price process $(S_t)_{t\in[0,T]}$ can be shown by changing the original measure $\mathbb{P}^0$ to the risk neutral measure $\mathbb{Q}$, defined by
    \begin{equation}
        \label{risk-neutral}
        \Bigl.\frac{d\mathbb{Q}}{d\mathbb{P}^{0}}\Bigr|_{\mathcal{F}^{0}_t} = \mathcal{E}\Bigl(-\int_0^\cdot \theta_s^\top dW^0_s \Bigr)_t,~~t\in[0,T],
    \end{equation}
    which is well-defined thanks to Assumption \ref{asm1}(iv).
\end{rem}

In this model, we consider a case in which agents can observe the stock prices but cannot identify their drifts and Brownian shocks independently. The available market information for agents is modelled by a filtration $\mathbb{G}^0$.

\begin{dfn}~\\
    $\mathbb{G}^0:=(\mathcal{G}_t^0)_{t\in[0,T]}$ is a complete and right-continuous filtration generated by the stock price process $(S_t)_{t\in[0,T]}$.
\end{dfn}
\begin{rem}
    Since $S_0\in\mathbb{R}^{d_0}_{++}$, $\mathcal{G}^0_0$ is trivial unlike $\mathcal{F}^0_0$.
\end{rem}
We set $\mathcal{G}^0:=\mathcal{G}^0_T$. It is obvious that $\mathcal{G}^0_t\subset\mathcal{F}^0_t$ for all $t\in[0,T]$. Define a process $\widetilde{W}^0$ by
\begin{equation}
    \begin{split}
        \label{tilde_W}
        \widetilde{W}^0_t := \int_0^t \sigma_r^{-1} \mathrm{diag}(S_r)^{-1}dS_r = W^0_t + \int_0^t \theta_s ds,~~~t\in[0,T].
    \end{split}
\end{equation}
We have the following property.
\begin{lem}~\\
    \label{G=FW}
    Let Assumption \ref{asm1} be in force. Moreover, let $\mathbb{F}^{\widetilde{W}^0}$ be a complete and right-continuous filtration generated by $(\widetilde{W}^0_t)_{t\in[0,T]}$. Then, we have $\mathbb{G}^0=\mathbb{F}^{\widetilde{W}^0}$.
\end{lem}
\noindent
\textbf{\textit{proof}}\\
Notice that the dynamics of $S$ is given by
\begin{equation}
    \begin{split}
        \label{S_with_tilde_W}
        S_t = S_0 + \int_0^t \mathrm{diag}(S_r)\sigma_r d\widetilde{W}_r^0,~~~t\in[0,T].
    \end{split}
\end{equation} 
Since $\sigma$ is bounded, the standard result for Lipschitz SDEs implies that \eqref{S_with_tilde_W} has a unique $\mathbb{F}^{\widetilde{W}^0}$-adapted solution. (cf. Remark \ref{well-posedness_S}.) This shows $\mathbb{G}^0\subset\mathbb{F}^{\widetilde{W}^0}$. Conversely, $\mathbb{G}^0\supset\mathbb{F}^{\widetilde{W}^0}$ is obvious by \eqref{tilde_W}. $\square$

By Girsanov's theorem, $\widetilde{W}^0$ is a $(\mathbb{G}^0,\mathbb{Q})$-Brownian motion, where $\mathbb{Q}$ is the risk-neutral measure defined in Remark \ref{risk-neutral}. We denote the expectation of the risk premium process $\theta$ conditionally on $\mathcal{G}^0_t$ by
\begin{equation}
    \begin{split}
        \label{theta_expected}
        \widehat\theta_t := \mathbb{E}[\theta_t|\mathcal{G}^0_t],~~~t\in[0,T].
    \end{split}
\end{equation}
Moreover, we introduce a process $\widehat{W}^0$ by
\begin{equation}
    \begin{split}
        \widehat{W}^0_t := \widetilde{W}^0_t - \int_0^t \widehat\theta_s ds =  W^0_t + \int_0^t (\theta_s-\widehat\theta_s) ds,~~~t\in[0,T],
    \end{split}
\end{equation}
which is called ``innovation process'' in the filtering theory. The dynamics of $S$ can be written as
\begin{equation}
    \begin{split}
        S_t&= S_0 + \int_0^t \mathrm{diag}(S_r)\sigma_r(\widehat\theta_rdr + d\widehat{W}^0_r),~~t\in[0,T].
    \end{split}
\end{equation}
The following property is well-known.
\begin{lem}
    Under Assumption \ref{asm1}, the process $\widehat{W}^0$ is a $(\mathbb{G}^0,\mathbb{P}^0)$-Brownian motion.
\end{lem}
\noindent
\textbf{\textit{proof}}\\
This is a consequence of L\'{e}vy's theorem. See, e.g. Pardoux \cite{pardoux_filtrage} [Proposition 2.2.7]. $\square$\\
\begin{rem}~\\
    Although the filtration $\mathbb{G}^{0}$ is larger than the augmented filtration generated by $\widehat{W}^0$ in general, we can show that every $(\mathbb{G}^{0},\mathbb{P}^{0})$-local martingale has a representation through a stochastic integral with respect to $\widehat{W}^0$. (See, e.g. Jeanblanc, Yor \& Chesney \cite{jeanblanc_mathematical_2009} [Proposition 1.7.7.1].)
\end{rem}

\subsection{Optimal investment problem with exponential utility}
Suppose there are countably infinitely many agents in the common financial market. The relevant probability spaces are defined as follows.\\

\noindent
(1) We denote by $(\Omega^i,\mathcal{F}^i,\mathbb{P}^i)$ ($i\in\mathbb{N}$) a complete probability space with a complete and right-continuous filtration $\mathbb{F}^i:=(\mathcal{F}^i_t)_{t\in[0,T]}$, generated by a $d$-dimensional standard Brownian motion $W^i:=(W^i_t)_{t\in[0,T]}$ and a $\sigma$-algebra $\sigma(\xi^i,x^i_0)$. $\mathcal{F}^i_0$ is the completion of $\sigma(\xi^i,x^i_0)$. 
Here, $\xi^i$ is an $\mathbb{R}$-valued random variable and $x^i_0$ is an $\mathbb{R}^d$-valued random variable. We set $\mathcal{F}^i:=\mathcal{F}^i_T$. \\

\noindent
(2) We denote by $(\Omega^{0,i},\mathcal{F}^{0,i},\mathbb{P}^{0,i})$ ($i\in\mathbb{N}$) a complete probability space with $\Omega^{0,i} := \Omega^0 \times \Omega^i$ and with $(\mathcal{F}^{0,i},\mathbb{P}^{0,i})$, which is the completion of $(\mathcal{F}^0 \otimes \mathcal{F}^i,\mathbb{P}^{0}\otimes \mathbb{P}^{i})$. Also, we define a $\sigma$-algebra $\mathcal{G}^{0,i}$ by the completion of $\mathcal{G}^0 \otimes \mathcal{F}^i$.
We denote by $\mathbb{F}^{0,i}:=(\mathcal{F}^{0,i}_t)_{t\in[0,T]}$ the complete and right-continuous augmentation of $(\mathcal{F}_t^0 \otimes \mathcal{F}_t^i)_{t\in[0,T]}$ and by $\mathbb{G}^{0,i}:=(\mathcal{G}^{0,i}_t)_{t\in[0,T]}$ the complete and right-continuous augmentation of $(\mathcal{G}_t^0 \otimes \mathcal{F}_t^i)_{t\in[0,T]}$.\\

\noindent
(3) We denote by $(\Omega,\mathcal{F},\mathbb{P})$ a complete probability space with $\Omega:=\prod_{i=0}^\infty\Omega^i$ and with $(\mathcal{F},\mathbb{P})$, which is the completion of $\Bigl(\bigotimes_{i=0}^\infty\mathcal{F}^i,\bigotimes_{i=0}^\infty\mathbb{P}^i\Bigr)$. The $\sigma$-algebra $\mathcal{G}$ is defined by the completion of  $\bigotimes_{i=1}^\infty\mathcal{F}^i\otimes\mathcal{G}^0$.
The filtration $\mathbb{F}:=(\mathcal{F}_t)_{t\in[0,T]}$ is the complete and right-continuous augmentation of $(\bigotimes_{i=0}^\infty\mathcal{F}^{i}_t)_{t\in[0,T]}$ and $\mathbb{G}:=(\mathcal{G}_t)_{t\in[0,T]}$ is the complete and right-continuous augmentation of $(\bigotimes_{i=1}^\infty\mathcal{F}^{i}_t\otimes \mathcal{G}^0_t)_{t\in[0,T]}$.\\

We denote by $\mathbb{E}[\cdot]$ the expectation with respect to $\mathbb{P}$ unless otherwise noted. In this paper, the heterogeneity of agents is characterized by $(W^i,\xi^i,x^i_0)_{i\in\mathbb{N}}$. The economy is modelled through an exponential quadratic Gaussian framework.
\begin{asm} ~\\
    \label{asm2}
    \textup{(i)} For each $i\in\mathbb{N}$, $\xi^i$ is an $\mathbb{R}$-valued, $\mathcal{F}^i_0$-measurable, and normally-distributed random variable representing agent-$i$'s initial wealth. $x^i_0$ is an $\mathbb{R}^d$-valued, $\mathcal{F}^i_0$-measurable, and normally-distributed random variable.
    
    \noindent
    \textup{(ii)} The random variables $\xi^i$ and $x_0^i$ are mutually independent for each $i\in\mathbb{N}$ and $(\xi^i,x_0^i)_{i\in\mathbb{N}}$ have the same distribution.

    \noindent
    \textup{(iii)} For each $i\in\mathbb{N}$, $(F^i)_{i\in\mathbb{N}}$ is an $\mathbb{R}$-valued and $\mathcal{G}_T^{0,i}$-measurable random variable, which represents the amount of liability of agent-$i$ at time $T$. Each $F^i$  is given by a quadratic form\footnote{The symbol $\langle \cdot,\cdot\rangle$ denotes the Euclidean inner product, i.e. $\langle x, y\rangle:=x^\top y$ for $x,y\in\mathbb{R}^n$.}
    \begin{equation}
        \begin{split}
        \label{EQG-F}
            F^i := \frac{1}{2} \langle A^{F}_{00}x^0_T,x^0_T\rangle + \frac{1}{2} \langle A^{F}_{11}x^i_T,x^i_T\rangle + \langle A^{F}_{10}x^0_T,x^i_T\rangle + \langle B^{F}_0,x^0_T\rangle + \langle B_1^{F},x^i_T\rangle + C^{F},
        \end{split}
    \end{equation}
    for $(A^{F}_{00}, A^{F}_{11}, A^{F}_{10},B^F_0,B^F_1,C^F)\in\mathbb{M}_{d_0}\times\mathbb{M}_{d}\times\mathbb{R}^{d\times d_0}\times\mathbb{R}^{d_0}\times\mathbb{R}^{d}\times\mathbb{R}$ and Gaussian factor processes $(x^0,x^i)\in\mathbb{L}^0(\mathbb{G}^0,\mathbb{R}^{d_0})\times\mathbb{L}^0(\mathbb{F}^i,\mathbb{R}^{d})$ defined by
    \begin{equation*}
        \begin{split}
            x_t^0 = x^0_0 -\int_0^t K_0(x^0_s - m_0)ds + \Sigma_0 \widehat{W}_t^0,~~~x_t^i = x_0^i -\int_0^t K(x^i_s - m)ds + \Sigma W_t^i,~~~t\in[0,T]
        \end{split}
    \end{equation*}
    for $x^0_0\in\mathbb{R}^{d_0}$, $(K_0,K)\in\mathbb{R}_{++}\times \mathbb{R}_{++}$, $(m_0,m)\in\mathbb{R}^{d_0}\times \mathbb{R}^{d}$, and $(\Sigma_0,\Sigma)\in\mathbb{R}^{d_0\times d_0}\times \mathbb{R}^{d\times d}$.

    \noindent
    \textup{(iv)} Each agent is a price taker; agent-$i$ must accept the prevailing prices as he/she has no market share to influence the price.
\end{asm}
\begin{rem}
    In this model, the agent-$i$'s liability $F^i$ is subject to both common noise and idiosyncratic noise. As an example of financial interpretation, suppose that the agents are financial firms and have derivative liability at time $T$. 
    In this case, $F^i$ denotes the total amount of payoff, which usually depends on the price of securities and the idiosyncratic information, such as the corporate size and the number of contracts or clients the agent-$i$ has.
\end{rem}

The trading strategy of agent-$i$ is denoted by an $\mathbb{R}^{d_0}$-valued, $\mathbb{G}^{0,i}$-progressively measurable process $\pi^i:=(\pi^i_t)_{t\in[0,T]}$. Each element of $\pi^i_t$ represents the amount of money invested in each stock at time $t$. The wealth process of agent-$i$ with strategy $\pi$ is denoted by $\mathcal{W}^{i,\pi}\in\mathbb{L}^0(\mathbb{G}^{0,i},\mathbb{R})$ and its dynamics is given by
\begin{equation}
    \begin{split}
        \mathcal{W}^{i,\pi}_t 
        &:= 
        \xi^i + \int_0^t \pi_r^\top \mathrm{diag}(S_r)^{-1}dS_r \\
        &= 
        \xi^i + \int_0^t \pi_s^\top\sigma_s\widehat\theta_sds + \int_0^t \pi_s^\top\sigma_s d\widehat{W}_s^0
    \end{split}
\end{equation}
for $t\in[0,T]$. The agents' problems are modelled on the probability space $(\Omega,\mathcal{G},\mathbb{P},\mathbb{G})$; for each $i\in\mathbb{N}$, agent-$i$ solves
\begin{equation}
    \begin{split}
        \sup_{\pi\in\mathbb{A}^i} \mathbb{E}\Bigl[-\exp\Bigl(-\gamma(\mathcal{W}^{i,\pi}_T-F^i)\Bigr) \Bigr]
    \end{split}
\end{equation}
subject to
\[
    \mathcal{W}^{i,\pi}_t =\xi^i + \int_0^t \pi_s^\top\sigma_s\widehat\theta_sds + \int_0^t \pi_s^\top\sigma_s d\widehat{W}_s^0,~~t\in[0,T].
\]
Here, $\gamma\in\mathbb{R}_{++}$ is the coefficient of absolute risk aversion and $\mathbb{A}^i$ is the admissible set for agent-$i$, whose definition is to be given. By writing $p_t:=\pi_t^\top\sigma_t$ for each $t\in[0,T]$, the problem can equivalently be written as
\begin{equation}
    \begin{split}
        \sup_{p\in\mathcal{A}^i} \mathbb{E}\Bigl[-\exp\Bigl(-\gamma(\mathcal{W}^{i,p}_T-F^i)\Bigr) \Bigr]
    \end{split}
\end{equation}
subject to
\[
    \mathcal{W}^{i,p}_t =\xi^i + \int_0^t p_s\widehat\theta_sds + \int_0^t p_s d\widehat{W}_s^0,~~~t\in[0,T],
\]
where the set $\mathcal{A}^i$ is defined by $\mathcal{A}^i:=\{p=\pi^\top\sigma;\pi\in\mathbb{A}^i\}$.\\

To deal with the optimal control problem, let us introduce a BSDE: for each $i\in\mathbb{N}$,
\begin{equation}
    \begin{split}
        \label{BSDE-optim}
        Y^i_t = F^i + \int_t^T \Bigl(-Z^{i,0}_s\widehat\theta_s - \frac{|\widehat\theta_s|^2}{2\gamma} + \frac{\gamma}{2}|Z^i_s|^2\Bigr)ds - \int_t^T Z^{i,0}_s d\widehat{W}^0_s - \int_t^T Z^{i}_s dW^i_s,~~~t\in[0,T].
    \end{split}
\end{equation}
Suppose that the BSDE \eqref{BSDE-optim} has a solution $(Y^i,Z^{i,0},Z^i)\in\mathbb{S}^2(\mathbb{G}^{0,i},\mathbb{R})\times\mathbb{H}^2(\mathbb{G}^{0,i},\mathbb{R}^{1\times d_0})\times\mathbb{H}^2(\mathbb{G}^{0,i},\mathbb{R}^{1\times d})$. 
Then, define a process $R^{i,p}\in\mathbb{L}^0(\mathbb{G}^{0,i},\mathbb{R})$ by
\begin{equation}
    \begin{split}
        R^{i,p}_t := -\exp\Bigl(-\gamma(\mathcal{W}^{i,p}_t-Y^i_t)\Bigr),~~~t\in[0,T],~~~i\in\mathbb{N}.
    \end{split}
\end{equation}

\begin{dfn} (Admissible space)\\
    The admissible space $\mathbb{A}^i$ is the set of trading strategies $\pi\in\mathbb{H}^2(\mathbb{P}^{0,i},\mathbb{G}^{0,i},\mathbb{R}^{d_0})$ such that a family $\{R^{i,p}_\tau;\tau\in\mathcal{T}(\mathbb{G}^{0,i})\}$ is uniformly integrable.
\end{dfn}

\begin{rem}~\\
    \textup{(i)} If the BSDE \eqref{BSDE-optim} has no solution, we set $\mathbb{A}^i=\emptyset$.\\
    \textup{(ii)} Since $F^i$ and $\theta$ are unbounded, the method of Kobylanski \cite{Kobylanski2000BackwardSD} cannot be applied to show the well-posedness of \eqref{BSDE-optim}. 
    The property of quadratic growth BSDE with unbounded generator and terminal value is studied by Briand \& Hu \cite{briandBSDEQuadraticGrowth2006, briandQuadraticBSDEsConvex2008}. In this paper, however, we do not delve into the general well-posedness result of \eqref{BSDE-optim} as we are going to search for a special solution in the exponential quadratic Gaussian framework.\\
    \textup{(iii)} The motivation of considering the BSDE \eqref{BSDE-optim} and the process $R^{i,p}$ is explained in Fujii \& Sekine \cite{fujiiMeanFieldEquilibriumPrice2023a} [Section 3.2]. This method is originally proposed by Hu, Imkeller \& M\"{u}ller \cite{huUtilityMaximizationIncomplete2005a}.
\end{rem}

\begin{thm}~\\
    \label{optimality}
    Let Assumption \ref{asm1} and \ref{asm2} be in force. 
    For each $i\in\mathbb{N}$, assume further that the BSDE \eqref{BSDE-optim} has a solution $(Y^i,Z^{i,0},Z^i)\in\mathbb{S}^2(\mathbb{G}^{0,i},\mathbb{R})\times\mathbb{H}^2(\mathbb{G}^{0,i},\mathbb{R}^{1\times d_0})\times\mathbb{H}^2(\mathbb{G}^{0,i},\mathbb{R}^{1\times d})$
    and that the process $p^{i,*}:=(p^{i,*}_t)_{t\in[0,T]}$ defined by
        \[
            p^{i,*}_t := Z^{i,0}_t + \frac{\widehat{\theta}^\top_t}{\gamma},~~~t\in[0,T]
        \]
        belongs to $\mathcal{A}^i$. Then, $p^{i,*}$ is an optimal strategy for agent-$i$.
\end{thm}
\noindent
\textbf{\textit{proof}}\\
To begin with, notice that $R^{i,p}_0=-e^{-\gamma(\xi^i-Y_0^i)}$ is independent of the control variable $p$. By Ito formula, we have
\begin{equation*}
    \begin{split}
        dR^{i,p}_t 
        &=
        R^{i,p}_t \Bigl(-\gamma d(\mathcal{W}^{i,p}_t-Y^i_t) + \frac{\gamma^2}{2}d\langle\mathcal{W}^{i,p}-Y^i\rangle_t\Bigr)\\
        &=
        \frac{\gamma^2}{2} R^{i,p}_t \Bigl|p_t- Z^{i,0}_t - \frac{\widehat{\theta}^\top_t}{\gamma} \Bigr|^2 dt + R^{i,p}_t (-\gamma(p_t-Z^{i,0}_t)d\widehat{W}^0_t + \gamma Z^i_t dW^i_t ).
    \end{split}
\end{equation*}
Then, for any $p\in\mathcal{A}^i$, we have
\[
    \frac{\gamma^2}{2} R^{i,p}_t \Bigl|p_t- Z^{i,0}_t - \frac{\widehat{\theta}^\top_t}{\gamma} \Bigr|^2 \leq 0,~~~dt\otimes \mathbb{P}\text{-}\mathrm{a.e.}
\]
Together with the definition of admissibility, this clearly implies that the process $R^{i,p}$ is a $(\mathbb{G}^{0,i},\mathbb{P}^{0,i})$-supermartingale for every $p\in\mathcal{A}^i$. Moreover, if we choose $p=p^{i,*}$, it holds that
\[
    dR^{i,p^{i,*}}_t = R^{i,p^{i,*}}_t (- \widehat{\theta}^\top_t d\widehat{W}^0_t + \gamma Z^i_t dW^i_t ).
\]
Having assumed $p^{i,*}\in\mathcal{A}^i$, we deduce that the process $R^{i,p^{i,*}}$ is a martingale. With these observations, we obtain a relation
\[
    \mathbb{E}\Bigl[-\exp\Bigl(-\gamma(\mathcal{W}^{i,p}_T-F^i)\Bigr) \Bigr] = \mathbb{E}[R^{i,p}_T]\leq \mathbb{E}\Bigl[-\exp\Bigl(-\gamma(\xi^i-Y_0^i)\Bigr) \Bigr] = \mathbb{E}[R^{i,p^{i,*}}_T]= \mathbb{E}\Bigl[-\exp\Bigl(-\gamma(\mathcal{W}^{i,p^{i,*}}_T-F^i)\Bigr) \Bigr]
\]
for any $p\in\mathcal{A}^i$. This indicates the optimality of $p^{i,*}$. $\square$

\section{Mean field equilibrium model under partial observation}
In this section, we construct the risk premium process endogenously under the asymptotic market clearing condition, whose definition is given below. 
Section 3.1 introduces a relevant mean field BSDE and finds its solution in a semi-analytical form by deriving an associated system of ordinary differential equations. 
Section 3.2 verifies that the solution obtained in Section 3.1 does indeed characterize the optimal strategy and the asymptotic market clearing. 
In Section 3.3, we derive the dynamics of the market risk premium process endogenously using the Kalman-Bucy filtering theory.

\subsection{The mean field BSDE}
\begin{dfn}~(Asymptotic market clearing condition)\\
    \label{market-clearing}
    The financial market satisfies the asymptotic market clearing condition (or the market clearing condition in the large population limit) if
    \begin{equation}
        \label{MC-eqn}
        \lim_{N\to\infty}\frac{1}{N}\sum_{i=1}^N \pi_t^{i,*} = 0,~~~dt\otimes \mathbb{P}\text{-}\mathrm{a.e.}
    \end{equation}
    holds. Here, $\pi_t^{i,*}$ denotes the optimal trading strategy of the agent-$i$.
\end{dfn}
From an economic perspective, this condition means that the excess demand (or supply) per capita converges to zero (in the sense of $dt\otimes \mathbb{P}$-almost everywhere) as the population of investors tends to infinity.
For each $i\in\mathbb{N}$, if all assumptions in Theorem \ref{optimality} hold, 
\[
    p^{i,*}_t :=(\pi_t^{i,*})^\top\sigma_t = Z^{i,0}_t + \frac{\widehat{\theta}^\top_t}{\gamma},~~~t\in[0,T]
\]
is an optimal strategy for agent-$i$. In this case, the asymptotic market clearing condition \eqref{MC-eqn} requires $\widehat{\theta}$ to satisfy
\[
    \lim_{N\to\infty}\frac{1}{N}\sum_{i=1}^N Z^{i,0}_t + \frac{\widehat\theta_t^\top}{\gamma} = 0,~~~dt\otimes \mathbb{P}\text{-}\mathrm{a.e.},
\] 
which is inconsistent with the assumption that $\widehat\theta$ is $\mathbb{G}^0$-adapted. Nevertheless, since the interactions among agents are symmetric and made only through $\widehat{\theta}$, the random variables $(Z^{i,0}_t)_{i\in\mathbb{N}}$ are expected to be exchangeable for each $t\in[0,T]$. 
Moreover, $\mathcal{F}^i_t$ and $\mathcal{F}^j_t$ being independent for $i\neq j$, we can expect, at least heuristically, that
\[
    \lim_{N\to\infty}\frac{1}{N}\sum_{i=1}^N Z^{i,0}_t = \mathbb{E}[Z^{i,0}_t|\mathcal{G}^0],~~~\mathbb{P}\text{-}\mathrm{a.s.}
\]
for each $t\in[0,T]$\footnote{
    For a $\mathbb{G}^{0,i}$-adapted process $X$, we have $\mathbb{E}[X_t|\mathcal{G}^0]=\mathbb{E}[X_t|\mathcal{G}^0_t],~\mathbb{P}^0$-a.s. for each $t\in[0,T]$ since $X_t$ is independent of $(\widehat{W}^0_s-\widehat{W}^0_t)_{s\in[t,T]}$.
}. 
For these reasons, we expect that the risk premium process $\theta\in\mathbb{H}^2(\mathbb{P}^0,\mathbb{F}^0,\mathbb{R}^{d_0})$ satisfying
\begin{equation}
    \label{hat_MRP_eqbm}
    \widehat{\theta}_t = -\gamma\mathbb{E}[Z^{i,0}_t|\mathcal{G}^0]^\top,~~~t\in[0,T]
\end{equation}
achieves the asymptotic market clearing condition. Such an observation motivates us to study the following mean field BSDE defined on $(\Omega^{0,i},\mathcal{G}^{0,i},\mathbb{P}^{0,i},\mathbb{G}^{0,i})$ for each $i\in\mathbb{N}$:
\begin{equation}
    \begin{split}
        \label{MF-BSDE1}
            Y^i_t&= F^i+ \int_t^T  \Bigl(\gamma Z^{i,0}_s\mathbb{E}[Z^{i,0}_s|\mathcal{G}^0]^{\top} - \frac{\gamma}{2}|\mathbb{E}[Z^{i,0}_s|\mathcal{G}^0]|^2 + \frac{\gamma}{2}|Z^i_s|^2 \Bigr)ds - \int_t^T Z^{i,0}_s d\widehat{W}^0_s - \int_t^T Z^i_s dW^i_s,~~~t\in[0,T].~~~~~
    \end{split}
  \end{equation}

The mean field BSDE \eqref{MF-BSDE1} can be shown to have a semi-analytical solution under certain assumptions. See also Fujii \& Sekine \cite{fujiiMeanFieldEquilibriumPrice2023a} [Section 4.1].

\begin{thm}~\\
    \label{EQG solution}
    Let Assumption \ref{asm1} and \ref{asm2} be in force. In addition, assume that the system of ordinary differential equations
    \begin{equation}
        \begin{split}
            \label{Riccati eqn}
            &\dot{A}_{00}(t) = -\gamma A_{00}(t)\Sigma_0\Sigma_0^\top A_{00}(t)  - \gamma A_{10}(t)^\top \Sigma\Sigma^\top A_{10}(t) + 2K_0  A_{00}(t), \\
            &\dot{A}_{11}(t) = -\gamma A_{11}(t) \Sigma\Sigma^\top A_{11}(t)   + 2K  A_{11}(t), \\
            &\dot{A}_{10}(t)  = -\gamma A_{10}(t)\Sigma_0\Sigma_0^\top A_{00}(t) - \gamma A_{11}(t) \Sigma\Sigma^\top A_{10}(t) + (K_0+K)A_{10}(t) ,\\
            &\dot{B}_0(t)=\Bigl(- \gamma A_{00}(t)\Sigma_0\Sigma_0^\top  + K_0\Bigr)B_{0}(t) - \gamma A_{10}(t)^\top\Sigma\Sigma^\top B_{1}(t) - K_0A_{00}(t)m_0 - KA_{10}(t)^\top m,\\
            &\dot{B}_1(t)= \Bigl(-\gamma A_{11}(t)\Sigma\Sigma^\top + K\Bigr)B_1(t) - \gamma \Bigl(A_{10}(t){\Sigma}_0{\Sigma}_0^\top A_{10}(t)^\top \mu_t^1 + A_{10}(t)\Sigma_0\Sigma_0^\top B_0(t)\Bigr) - KA_{11}(t)m - K_0A_{10}(t)m_0, \\
            &\dot{C}(t)=  - \frac{\gamma}{2}\langle \Sigma_0^\top B_0(t),\Sigma_0^\top B_0(t) \rangle - \frac{\gamma}{2} \langle \Sigma^\top B_1(t),\Sigma^\top B_1(t) \rangle - \langle K_0 B_0(t) ,m_0 \rangle - \langle K B_1(t) ,m \rangle \\
            &~~~~~~~~~~~~~+ \frac{\gamma}{2}\langle A_{10}(t){\Sigma}_0{\Sigma}_0^\top A_{10}(t)^\top \mu_t^1,\mu_t^1 \rangle - \frac{1}{2}\mathrm{tr}[A_{00}(t)\Sigma_0\Sigma_0^\top] - \frac{1}{2}\mathrm{tr}[A_{11}(t)\Sigma\Sigma^\top],\\
            &A_{00}(T)=A_{00}^F,~~ A_{11}(T)=A_{11}^F, ~~A_{10}(T)=A_{10}^F, ~~B_0(T)=B_0^F, ~~B_1(T)=B_1^F, ~~C(T)=C^F
        \end{split}
    \end{equation}
    for $t\in[0,T]$ has a global solution $(A_{00},A_{11},A_{10},B_0,B_1,C)\in\mathcal{C}^1([0,T];\mathbb{M}_{d_0})\times\mathcal{C}^1([0,T];\mathbb{M}_{d})\times\mathcal{C}^1([0,T];\mathbb{R}^{d\times d_0})\times\mathcal{C}^1([0,T];\mathbb{R}^{d_0})\times\mathcal{C}^1([0,T];\mathbb{R}^{d})\times\mathcal{C}^1([0,T];\mathbb{R})$. Here, $\mu^1_t:=\mathbb{E}[x^1_t] = \mathbb{E}[x_0^1]e^{-K t}+m(1-e^{-K t})$ for $t\in[0,T]$.
    Then, for each $i\in\mathbb{N}$, the processes $(Y^i,Z^{i,0},Z^i)\in\mathbb{S}^2(\mathbb{P}^{0,i},\mathbb{G}^{0,i},\mathbb{R}) \times \mathbb{H}^2(\mathbb{P}^{0,i},\mathbb{G}^{0,i},\mathbb{R}^{1\times d_0}) \times \mathbb{H}^2(\mathbb{P}^{0,i},\mathbb{G}^{0,i},\mathbb{R}^{1\times d})$, defined by
    \begin{equation}
        \begin{split}
            \label{EQG solution YZ}
            &Y^i_t  := \frac{1}{2} \langle A_{00}(t)x^0_t,x^0_t\rangle + \frac{1}{2} \langle A_{11}(t)x^i_t,x^i_t\rangle + \langle A_{10}(t)x^0_t,x^i_t\rangle + \langle B_0(t),x^0_t\rangle + \langle B_1(t),x^i_t\rangle + C(t), \\
            &Z_t^{i,0} := \Bigl\{\Sigma_0^\top (A_{00}(t)x^0_t + A_{10}(t)^\top x^i_t + B_0(t))\Bigr\}^\top,~~~Z_t^i := \Bigl\{\Sigma^\top (A_{10}(t)x^0_t + A_{11}(t)x^i_t + B_1(t))\Bigr\}^\top
        \end{split}
    \end{equation}
    for $t\in[0,T]$, solve the mean field BSDE \eqref{MF-BSDE1}.
\end{thm}

\noindent
\textbf{\textit{proof}}\\
By the terminal condition of \eqref{Riccati eqn} and Assumption \ref{asm2} (iii), it follows that $Y^i_T = F^i$. Applying Ito formula to \eqref{EQG solution YZ}, we have
\begin{equation*}
    \begin{split}
        \label{Y-ito}
        dY^i_t 
        &=
        \Bigl\{\Bigl\langle \Bigl(\frac{1}{2} \dot{A}_{00}(t) - K_0 A_{00}(t)\Bigr)x^0_t,x^0_t\Bigr\rangle + \Bigl\langle \Bigl(\frac{1}{2} \dot{A}_{11}(t) - K A_{11}(t)\Bigr)x^i_t,x^i_t\Bigr\rangle + \Bigl\langle \Bigl( \dot{A}_{10}(t) - (K_0+K)A_{10}(t)\Bigr)x^0_t,x^i_t\Bigr\rangle \Bigr.~~~~ \\
        &~~~+ \langle \dot{B}_0(t)-K_0B_0(t) + K_0A_{00}(t)m_0 + K A_{10}(t)^\top m,x^0_t\rangle + \langle \dot{B}_1(t)-K B_1(t) + K A_{11}(t)m + K_0A_{10}(t)m_0,x^i_t\rangle \\
        &~~~+ \Bigl.\dot{C}(t) + \langle K_0 B_0(t) ,m_0 \rangle + \langle K B_1(t) ,m \rangle +\frac{1}{2}\mathrm{tr}[A_{00}(t)\Sigma_0\Sigma_0^\top] + \frac{1}{2}\mathrm{tr}[A_{11}(t)\Sigma\Sigma^\top] \Bigr\}dt \\
        &~~~+   \langle\Sigma_0^\top (A_{00}(t)x^0_t + A_{10}(t)^\top x^i_t + B_0(t)) ,d\widehat{W}^0_t \rangle + \langle \Sigma^\top (A_{10}(t)x^0_t + A_{11}(t)x^i_t + B_1(t)) , dW^i_t \rangle.\\
        &=
        \Bigl\{\Bigl\langle \Bigl(\frac{1}{2} \dot{A}_{00}(t) - K_0 A_{00}(t)\Bigr)x^0_t,x^0_t\Bigr\rangle + \Bigl\langle \Bigl(\frac{1}{2} \dot{A}_{11}(t) - K A_{11}(t)\Bigr)x^i_t,x^i_t\Bigr\rangle + \Bigl\langle \Bigl( \dot{A}_{10}(t) - (K_0+K)A_{10}(t)\Bigr)x^0_t,x^i_t\Bigr\rangle \Bigr.~~~~ \\
        &~~~+ \langle \dot{B}_0(t)-K_0B_0(t) + K_0A_{00}(t)m_0 + K A_{10}(t)^\top m,x^0_t\rangle + \langle \dot{B}_1(t)-K B_1(t) + K A_{11}(t)m + K_0A_{10}(t)m_0,x^i_t\rangle \\
        &~~~+ \Bigl.\dot{C}(t) + \langle K_0 B_0(t) ,m_0 \rangle + \langle K B_1(t) ,m \rangle +\frac{1}{2}\mathrm{tr}[A_{00}(t)\Sigma_0\Sigma_0^\top] + \frac{1}{2}\mathrm{tr}[A_{11}(t)\Sigma\Sigma^\top] \Bigr\}dt \\
        &~~~+   Z^{i,0}_t d\widehat{W}^0_t + Z^i_t dW^i_t
    \end{split}
\end{equation*}
for $t\in[0,T]$. By the ODE \eqref{Riccati eqn}, it holds that
\begin{equation}
    \begin{split}
        &\Bigl\langle \Bigl(\frac{1}{2} \dot{A}_{00}(t) - K_0 A_{00}(t)\Bigr)x^0_t,x^0_t\Bigr\rangle + \Bigl\langle \Bigl(\frac{1}{2} \dot{A}_{11}(t) - K A_{11}(t)\Bigr)x^i_t,x^i_t\Bigr\rangle + \Bigl\langle \Bigl( \dot{A}_{10}(t) - (K_0+K)A_{10}(t)\Bigr)x^0_t,x^i_t\Bigr\rangle \Bigr.~~~~ \\
        &~~~+ \langle \dot{B}_0(t)-K_0B_0(t) + K_0A_{00}(t)m_0 + K A_{10}(t)^\top m,x^0_t\rangle + \langle \dot{B}_1(t)-K B_1(t) + K A_{11}(t)m + K_0A_{10}(t)m_0,x^i_t\rangle \\
        &~~~+ \Bigl.\dot{C}(t) + \langle K_0 B_0(t) ,m_0 \rangle + \langle K B_1(t) ,m \rangle +\frac{1}{2}\mathrm{tr}[A_{00}(t)\Sigma_0\Sigma_0^\top] + \frac{1}{2}\mathrm{tr}[A_{11}(t)\Sigma\Sigma^\top]  \\
        &=
        -\Bigl\langle \frac{\gamma}{2}\Bigl(A_{00}(t)\Sigma_0\Sigma_0^\top A_{00}(t) + A_{10}(t)^\top \Sigma\Sigma^\top A_{10}(t) \Bigr) x_t^0,x_t^0 \Bigr\rangle - \Bigl\langle \frac{\gamma}{2}A_{11}(t) \Sigma\Sigma^\top A_{11}(t) x_t^i ,x_t^i \Bigr\rangle \\
        &~~~-\Bigl\langle \gamma (A_{10}(t)\Sigma_0\Sigma_0^\top A_{00}(t) + A_{11}(t) \Sigma\Sigma^\top A_{10}(t) )x_t^0,x_t^i \Bigr\rangle \\
        &~~~-\Bigl\langle \gamma (A_{00}(t)\Sigma_0\Sigma_0^\top B_{0}(t) + A_{10}(t)^\top\Sigma\Sigma^\top B_{1}(t))  ,x_t^0 \Bigr\rangle \\
        &~~~-\Bigl\langle \gamma (A_{10}(t){\Sigma}_0{\Sigma}_0^\top A_{10}(t)^\top \mu_t^1 + A_{10}(t)\Sigma_0\Sigma_0^\top B_0(t) + A_{11}(t)\Sigma\Sigma^\top B_1(t)) ,x_t^i \Bigr\rangle\\
        &~~~+ \frac{\gamma}{2}\langle A_{10}(t){\Sigma}_0{\Sigma}_0^\top A_{10}(t)^\top \mu_t^1,\mu_t^1 \rangle - \frac{\gamma}{2}\langle \Sigma_0^\top B_0(t),\Sigma_0^\top B_0(t) \rangle - \frac{\gamma}{2}\langle \Sigma^\top B_1(t),\Sigma^\top B_1(t) \rangle\\
        &=
        -\gamma Z^{i,0}_t\mathbb{E}[Z^{i,0}_t|\mathcal{G}^0]^{\top} + \frac{\gamma}{2}|\mathbb{E}[Z^{i,0}_t|\mathcal{G}^0]|^2 - \frac{\gamma}{2}|Z^i_t|^2
    \end{split}
\end{equation}
for $t\in[0,T]$. Here, we used
\begin{equation}
    \begin{split}
        \mathbb{E}[Z^{i,0}_t|\mathcal{G}^0] = \Bigl\{\Sigma_0^\top (A_{00}(t)x^0_t + A_{10}(t)^\top \mu^1_t + B_0(t))\Bigr\}^\top,~~~t\in[0,T]
    \end{split}
\end{equation}
in the last equality, since $x^0_t$ is $\mathcal{G}^0$-measurable and $x^i_t$ ($i\in\mathbb{N}$) is independent of $\mathcal{G}^0$ for every $t\in[0,T]$.\footnote{By the construction of the probability spaces, $W^i$ ($i\in\mathbb{N}$) is independent of ${W}^0$ on $(\Omega, \mathcal{F}, \mathbb{P})$.} These observations show
\begin{equation}
    \begin{split}
        dY^i_t 
        &=
        -\Bigl(\gamma Z^{i,0}_t\mathbb{E}[Z^{i,0}_t|\mathcal{G}^0]^{\top} - \frac{\gamma}{2}|\mathbb{E}[Z^{i,0}_t|\mathcal{G}^0]|^2 + \frac{\gamma}{2}|Z^i_t|^2 \Bigr)dt + Z^{i,0}_t d\widehat{W}^0_t + Z^i_t dW^i_t,~~~t\in[0,T],~~~Y^i_T = F^i,~~~\mathbb{P}^{0,i}\text{-}\mathrm{a.s.}
    \end{split}
\end{equation}
i.e. $(Y^i,Z^{i,0},Z^i)$ solve the mean field BSDE \eqref{MF-BSDE1}. $\square$

\begin{rem}~\\
    \label{rem1}
    \textup{(i)} Notice that, for each $i\in\mathbb{N}$, if $A^F_{10}=0$, namely $F^i$ has no cross-term of $x^i_T$ and $x^0_T$, we can write $F^i = \widetilde{F}^0 + \widetilde{F}^i$, where 
    \begin{equation}
        \begin{split}
        \widetilde{F}^0 := \frac{1}{2} \langle A^{F}_{00}x^0_T,x^0_T\rangle + \langle B^{F}_0,x^0_T\rangle  + C^{F},~~~
        \widetilde{F}^i :=\frac{1}{2} \langle A^{F}_{11}x^i_T,x^i_T\rangle  + \langle B_1^{F},x^i_T\rangle.
        \end{split}
    \end{equation}  
    It is clear that $\widetilde{F}^0$ (resp. $\widetilde{F}^i$) is an $\mathcal{G}^0_T$-measurable (resp. $\mathcal{F}^i_T$-measurable) random variable. In this case, we can find a solution to the mean field BSDE \eqref{MF-BSDE1} by considering these two non-mean field BSDEs:
    \begin{equation}
        \begin{split}
            \label{separated-BSDE}
            \widetilde{Y}^0_t &= \widetilde{F}^0 + \int_t^T \frac{\gamma}{2}|\widetilde{Z}^0_s|^2ds - \int_t^T \widetilde{Z}^0_s d\widehat{W}^0_s,~~~t\in[0,T],\\
            \widetilde{Y}^i_t &= \widetilde{F}^i + \int_t^T \frac{\gamma}{2}|\widetilde{Z}^i_s|^2ds - \int_t^T \widetilde{Z}^i_s dW^i_s,~~~t\in[0,T].
        \end{split}
    \end{equation}
    Indeed, if the BSDEs $\eqref{separated-BSDE}$ have solutions $(\widetilde{Y}^0,\widetilde{Z}^0)$ and $(\widetilde{Y}^i,\widetilde{Z}^i)$, we deduce that $\widetilde{Y}^0$ also solves
    \[
        \widetilde{Y}^0_t = \widetilde{F}^0 + \int_t^T \Bigl(\gamma Z^{0}_s\mathbb{E}[Z^{0}_s|\mathcal{G}^0]^{\top} - \frac{\gamma}{2}|\mathbb{E}[Z^{0}_s|\mathcal{G}^0]|^2\Bigr)ds - \int_t^T \widetilde{Z}^0_s d\widehat{W}^0_s,~~~t\in[0,T]
    \]
    since $\mathbb{E}[Z^{0}_t|\mathcal{G}^0] = Z^{0}_t$. Then, it is clear that $(\widetilde{Y}^0 + \widetilde{Y}^i, \widetilde{Z}^0, \widetilde{Z}^i)$ solves the mean field BSDE \eqref{MF-BSDE1}. See also Fujii \& Sekine \cite{fujiiMeanFieldEquilibriumPrice2023a} [Section 4.3].\\
    \textup{(ii)} If $\exp(\widetilde{F}^0)$ and $\exp(\widetilde{F}^i)$ are integrable, $\eqref{separated-BSDE}$ have closed-form solutions: 
    \[
        \widetilde{Y}^0_t=\log\mathbb{E}[\exp(\widetilde{F}^0)|\mathcal{G}^0_t],~~~\widetilde{Y}^i_t=\log\mathbb{E}[\exp(\widetilde{F}^i)|\mathcal{F}^i_t],~~~t\in[0,T].
    \]
    \textup{(iii)} For instance, $F^i$ with $A^F_{10}=0$ can be interpreted as a liability that is additively separated into the performance of a benchmark portfolio $\widetilde{F}^0$ quoted in the market, and an additional gain $\widetilde{F}^i$ required by the manager or clients of agent-$i$.\\  
    \textup{(iv)} By the local Lipschitz property, the solution of \eqref{Riccati eqn} is locally unique if exists. As a result, \eqref{EQG solution YZ} is the unique solution to the BSDE \eqref{MF-BSDE1} among those of the form \eqref{EQG solution YZ}.\\
    \textup{(v)} In this model, agents are assumed to be homogeneous in the risk-aversion parameter in order to simplify the mathematical analysis. 
    We may possibly allow heterogeneity in the risk-aversion parameter, namely agent-$i$'s risk-aversion parameter is expressed by $\mathcal{F}^i_0$-measurable positive random variable $\gamma^i$ for each $i\in\mathbb{N}$
    \footnote{The definition of the filtered probalibity space $(\Omega^i,\mathcal{F}^i,\mathbb{P}^i,\mathbb{F}^i)$ should be modified to make $\gamma^i$ measurable; $\mathcal{F}^i_0$ is set to be a completion of $\sigma(\xi^i,x_0^i,\gamma^i)$. 
    We also assume that $0<\underline{\gamma}\leq\gamma^i\leq \overline{\gamma}$ ($i\in\mathbb{N}$) for some constants $0<\underline{\gamma}\leq\overline{\gamma}$ and that $(\gamma^i)_{i\in\mathbb{N}}$ are i.i.d. on $(\Omega,\mathcal{F},\mathbb{P})$. See \cite{fujiiMeanFieldEquilibriumPrice2023a} for the general settings.}.
    In such a case, however, we need to consider a system of mean field type ODEs, whose well-posedness is much more difficult to prove than \eqref{Riccati eqn}. See also Fujii \& Sekine \cite{fujiiMeanFieldEquilibriumAsset2024} [Section 4.2].
\end{rem}

\subsection{Optimal control and asymptotic market clearing condition}
Suppose that the equation \eqref{Riccati eqn} has a global solution $(A_{00},A_{11},A_{10},B_0,B_1,C)\in\mathcal{C}^1([0,T];\mathbb{M}_{d_0})\times\mathcal{C}^1([0,T];\mathbb{M}_{d})\times\mathcal{C}^1([0,T];\mathbb{R}^{d\times d_0})\times\mathcal{C}^1([0,T];\mathbb{R}^{d_0})\times\mathcal{C}^1([0,T];\mathbb{R}^{d})\times\mathcal{C}^1([0,T];\mathbb{R})$
and define the processes $(Y^i,Z^{i,0},Z^i)$ by \eqref{EQG solution YZ}.
From \eqref{hat_MRP_eqbm}, if the market risk premium process $\theta$ satisfies
\begin{equation}
    \label{eqbm-theta}
    \widehat\theta_t := \mathbb{E}[\theta_t|\mathcal{G}^0_t] = -\gamma\mathbb{E}[Z^{i,0}_t|\mathcal{G}^0]^\top = -\gamma {\Sigma}_0^\top \Bigl(A_{00}(t)x^0_t + A_{10}(t)^\top \mu^1_t + B_0(t) \Bigr),~~~t\in[0,T],
\end{equation}
we expect that the asymptotic market clearing condition is satisfied with strategies
\begin{equation}
    \begin{split}
        \label{EQG-optimal}
        p^{i,*}_t &:= (\pi^{i,*}_t)^\top\sigma_t := Z^{i,0}_t + \frac{\widehat\theta_t^\top}{\gamma},~~~t\in[0,T],~~~i\in\mathbb{N}.
    \end{split}
\end{equation}
The following theorem proves this observation under additional assumptions.
\begin{thm} ~\\
    \label{EQG-verification}
    Let Assumptions \ref{asm1} and \ref{asm2} be in force. Assume further that the equation \eqref{Riccati eqn} has a global solution $(A_{00},A_{11},A_{10},B_0,B_1,C)\in\mathcal{C}^1([0,T];\mathbb{M}_{d_0})\times\mathcal{C}^1([0,T];\mathbb{M}_{d})\times\mathcal{C}^1([0,T];\mathbb{R}^{d\times d_0})\times\mathcal{C}^1([0,T];\mathbb{R}^{d_0})\times\mathcal{C}^1([0,T];\mathbb{R}^{d})\times\mathcal{C}^1([0,T];\mathbb{R})$
    and that $\mathrm{Var}(x^1_0)^{-1}-\gamma A_{11}(0)$ is a positive definite matrix. Then, if the market risk premium process $\theta$ satisfies \eqref{eqbm-theta}, the following statements hold. \\
    (1) For each $i\in\mathbb{N}$, the process $p^{i,*}$, defined by \eqref{EQG-optimal}, is an optimal strategy for agent-$i$. \\
    (2) The asymptotic market clearing condition \eqref{MC-eqn} is satisfied as long as each agent adopts \eqref{EQG-optimal} as his/her optimal strategy.\\
    Here, $\mathrm{Var}(x^1_0)$ is the covariance matrix of $x^1_0$, i.e. $\mathrm{Var}(x^1_0):=\mathbb{E}[(x^1_0-\mathbb{E}[x^1_0])(x^1_0-\mathbb{E}[x^1_0])^\top]$ and the processes $(Y^i,Z^{i,0},Z^i)\in\mathbb{S}^2(\mathbb{P}^{0,i},\mathbb{G}^{0,i},\mathbb{R}) \times \mathbb{H}^2(\mathbb{P}^{0,i},\mathbb{G}^{0,i},\mathbb{R}^{1\times d_0}) \times \mathbb{H}^2(\mathbb{P}^{0,i},\mathbb{G}^{0,i},\mathbb{R}^{1\times d})$ are given by \eqref{EQG solution YZ}. 
\end{thm}

\noindent
\textbf{\textit{proof}}\\
For (1), it suffices to show $p^{i,*}\in\mathcal{A}^i$ by Theorem \ref{optimality} and \ref{EQG solution}.
In this proof, we use $\widetilde{C}>0$ as a general constant, whose value may change line by line. \par
Recall that $(x^i_0)_{i\in\mathbb{N}}$ are Gaussian random variables and are independently and identically distributed (i.i.d.) on $(\Omega,\mathcal{G},\mathbb{P})$. If $\mathrm{Var}(x^1_0)^{-1}-\gamma A_{11}(0)$ is positive definite, we have
\begin{equation}
    \begin{split}
    \mathbb{E}[e^{\gamma Y^i_0}]
    &=
    \mathbb{E}\Bigl[\exp\Bigl(\frac{\gamma}{2} \langle A_{00}(0)x^0_0,x^0_0\rangle + \frac{\gamma}{2} \langle A_{11}(0)x^i_0,x^i_0\rangle + \gamma\langle A_{10}(0)x^0_0,x^i_0\rangle + \gamma\langle B_0(0),x^0_0\rangle + \gamma\langle B_1(0),x^i_0\rangle + \gamma C(0)\Bigr)\Bigr]\\
    &=
    \widetilde{C}\mathbb{E}\Bigl[\exp\Bigl(\frac{\gamma}{2} \langle A_{11}(0)x^i_0,x^i_0\rangle + \gamma\langle A_{10}(0)x^0_0 + B_1(0),x^i_0\rangle\Bigr)\Bigr]\\
    &=
    \widetilde{C} \int_{\mathbb{R}^{d_0}} \exp\Bigl(-\frac{1}{2}(\bm{x}-\mu^1_0)^\top \mathrm{Var}(x^1_0)^{-1} (\bm{x}-\mu^1_0) + \frac{\gamma}{2} \bm{x}^\top A_{11}(0)\bm{x} + \gamma\langle A_{10}(0)x^0_0 + B_1(0),\bm{x}\rangle\Bigr) d\bm{x}\\
    &\leq
    \widetilde{C} \int_{\mathbb{R}^{d_0}} \exp\Bigl(-\frac{1}{2}\bm{x}^\top (\mathrm{Var}(x^1_0)^{-1}-\gamma A_{11}(0))\bm{x} + \gamma\langle A_{10}(0)x^0_0 + B_1(0) + \mathrm{Var}(x^1_0)^{-1}\mu^1_0,\bm{x}\rangle \Bigr) d\bm{x}\\
    &<
    \infty.
    \end{split}
\end{equation}
Since $\xi^i$ is independent of $x^i_0$ and normally distributed, this observation implies
\[
    \mathbb{E}[e^{-\gamma (\xi^i-Y^i_0)}] = \mathbb{E}[e^{-\gamma \xi^i}] \mathbb{E}[e^{\gamma Y^i_0}]  < \infty.
\]

We have
\[
    R^{i,p^{i,*}}_t = -\exp\Bigl(-\gamma(\xi^i-Y_0^i)\Bigr)\mathcal{E}\Bigl(-\int_0^\cdot \widehat\theta^\top_s d\widehat{W}_s^0 + \int_0^\cdot \gamma Z^i_s dW^i_s\Bigr)_t,~~~t\in[0,T]
\]
by the definition of $R^{i,p}$ and $p^{i,*}$. Define a process $V^i\in\mathbb{L}^0(\mathbb{G}^{0,i},\mathbb{R}_{++})$ by
\[
    V^i_t := \mathcal{E}\Bigl(-\int_0^\cdot \widehat\theta^\top_s d\widehat{W}_s^0 + \int_0^\cdot \gamma Z^i_s dW^i_s\Bigr)_t,~~~t\in[0,T].
\]
By writing $\Theta^i := (-\widehat\theta^\top, \gamma Z^i) \in\mathbb{S}^2(\mathbb{P}^{0,i},\mathbb{G}^{0,i},\mathbb{R}^{1\times (d_0 + d)})$ and $\bm{W}^{0,i}:= \begin{pmatrix}
    \widehat{W}^0 \\
    W^i \\
    \end{pmatrix}
    $, $V^i$ can be written as
\[
    V^i_t = \mathcal{E}\Bigl(\int_0^\cdot \Theta^i_s d\bm{W}^{0,i}_s\Bigr)_t,~~~t\in[0,T].
\]
We set $\bm{x}^{0,i}:= \begin{pmatrix}
    x^0 \\
    x^i \\
    \end{pmatrix}
    $. Then, $\bm{x}^{0,i}$ follows the dynamics
\[
    d\bm{x}^{0,i}_t = -\bm{K} \Bigl(\bm{x}^{0,i}_t - \bm{m}\Bigr) dt + \bm{\Sigma} d\bm{W}^{0,i}_s,
\]
where
\[
    \bm{K} := \begin{pmatrix} K_0 I_{d_0} & 0 \\ 0 & K I_{d} \\ \end{pmatrix},~~~\bm{m}:=\begin{pmatrix} m_0 \\ m \\ \end{pmatrix},~~~\bm{\Sigma}:= \begin{pmatrix} \Sigma_0 & 0 \\ 0 & \Sigma \\ \end{pmatrix}.
\]
Note that $\bm{x}^{0,i}_0\in\mathbb{L}^2(\mathbb{P}^{0,i},\mathcal{G}^{0,i}_0,\mathbb{R}^{d_0 + d})$, $|\Theta^i_t|^2\leq \widetilde{C}(|\widehat\theta_t|^2 + |Z^i_t|^2) \leq \widetilde{C}(1+|\bm{x}^{0,i}_t|^2)$ for all $t\in[0,T]$ and that $\bm{W}^i$ is a $(d_0 + d)$-dimensional standard $(\mathbb{G}^{0,i},\mathbb{P}^{0,i})$-Brownian motion. 
Then, by Bain \& Crisan \cite{bain_fundamentals_2009} [Exercise 3.11], $V^i$ is a $(\mathbb{G}^{0,i},\mathbb{P}^{0,i})$-martingale. It is now easy to see that
\[
    \mathbb{E}[|R^{i,p^{i,*}}_t|] = \mathbb{E}[e^{-\gamma (\xi^i-Y^i_0)}V^i_t] = \mathbb{E}[e^{-\gamma (\xi^i-Y^i_0)} \mathbb{E}[V^i_t|\mathcal{G}_0^{0,i}]] = \mathbb{E}[e^{-\gamma (\xi^i-Y^i_0)}] < \infty,~~~t\in[0,T],
\]
and that, for all $0\leq s\leq t\leq T$,
\[
    \mathbb{E}[R^{i,p^{i,*}}_t|\mathcal{G}^{0,i}_s]=  -e^{-\gamma (\xi^i-Y^i_0)}\mathbb{E}[V^i_t|\mathcal{G}^{0,i}_s] = -e^{-\gamma (\xi^i-Y^i_0)}V^i_s = R^{i,p^{i,*}}_s,~~~\mathbb{P}^{0,i}\text{-}\mathrm{a.s.}
\]
This clearly shows that $R^{i,p^{i,*}}$ is a martingale. By the optional sampling theorem and $\mathbb{E}[|R^{i,p^{i,*}}_T|]<\infty$, we conclude that the family $\{R^{i,p^{i,*}}_\tau;\tau\in\mathcal{T}(\mathbb{G}^{0,i})\}$ is uniformly integrable, i.e. $p^{i,*}\in\mathcal{A}^i$. 

We now verify (2). Notice that $\pi^{i,*}$ can be written as
\[
    \pi^{i,*}_t = (\sigma_t^\top)^{-1} (p_t^{i,*})^\top  = (\sigma_t^\top)^{-1} \Sigma_0^\top A_{10}(t)^\top (x^i_t-\mu^1_t),~~~t\in[0,T].
\]
Since, for each $t\in[0,T]$, $(x^i_t)_{i\in\mathbb{N}}$ are i.i.d. and $\mathbb{E}[x^i_t]=\mu^1_t$ for all $i\in\mathbb{N}$, we have
\[
    \mathbb{E}\int_0^T \left|\frac{1}{N} \sum_{i=1}^N \pi_t^{i,*}\right|^2 dt \leq \widetilde{C} \mathbb{E}\int_0^T \left|\frac{1}{N} \sum_{i=1}^N (x^i_t-\mu^1_t)\right|^2 dt \leq \frac{\widetilde{C}}{N^2} \sum_{i=1}^N \mathbb{E}\int_0^T \left|x^i_t-\mu^1_t\right|^2 dt \leq \frac{\widetilde{C}}{N}\to 0, ~~(N\to\infty),
\]
which implies \eqref{MC-eqn}. $\square$

\begin{rem}
    If the matrix $A^F_{11}$ is negative semidefinite, the ODE
    \[
        \dot{A}_{11}(t) = -\gamma A_{11}(t) \Sigma\Sigma^\top A_{11}(t)   + 2K  A_{11}(t),~~~t\in[0,T],~~~A_{11}(T)=A^F_{11}
    \]
    has a unique solution on $[0,T]$ for any $T>0$ (See, e.g. \cite{mat_riccati} [Theorem 8]), and the solution $A^F_{11}(t)$ is negative semidefinite for all $t\in[0,T]$ (See, e.g. \cite{mat_riccati} [Theorem 9].) In such a case, the condition that $\mathrm{Var}(x^1_0)^{-1}-\gamma A_{11}(0)$ is positive definite is satisfied.
\end{rem}

\subsection{Market risk premium process}
In this section, we assume that the risk premium process $\theta$ follows a linear Gaussian dynamics on $(\mathbb{P}^0,\mathbb{F}^0)$. Using the Kalman-Bucy filtering theory, we construct a semi-explicit formulation of the risk premium process.
\begin{asm}~\\
    \label{asm_kalman}
    \textup{(i)} The market risk premium process $\theta$ follows
                \begin{equation}
                    \begin{split}
                        \label{LG_MRP}
                        \theta_t = \theta_0 + \int_0^t (\alpha_s\theta_s + \beta_s)ds + \int_0^t \zeta_s dW^0_s + \int_0^t \eta_s dB^0_s,~~~t\in[0,T],
                    \end{split}
                \end{equation}
                for $\alpha\in\mathcal{C}([0,T];\mathbb{R}^{d_0\times d_0})$, $\beta\in\mathcal{C}([0,T];\mathbb{R}^{d_0})$, $\zeta\in\mathcal{C}^1([0,T];\mathbb{M}_{d_0})$ and $\eta\in\mathcal{C}([0,T];\mathbb{R}^{d_0\times k})$. The initial condition $\theta_0\in\mathbb{L}^2(\mathbb{P}^0,\mathcal{F}^0_0,\mathbb{R}^{d_0})$ is normally distributed: $\theta_0\sim N(m,v)$ for $(m,v)\in\mathbb{R}^{d_0}\times \mathbb{M}_{d_0}$. \\
    \textup{(ii)} $\Sigma_0$ is invertible.\\
    \textup{(iii)} The system of ordinary differential equations \eqref{Riccati eqn} has a global solution $(A_{00},A_{11},A_{10},B_0,B_1,C)\in\mathcal{C}^1([0,T];\mathbb{M}_{d_0})\times\mathcal{C}^1([0,T];\mathbb{M}_{d})\times\mathcal{C}^1([0,T];\mathbb{R}^{d\times d_0})\times\mathcal{C}^1([0,T];\mathbb{R}^{d_0})\times\mathcal{C}^1([0,T];\mathbb{R}^{d})\times\mathcal{C}^1([0,T];\mathbb{R})$ and $A_{00}(t)$ is invertible for all $t\in[0,T]$.\\
    \textup{(iv)} $\mathrm{Var}(x^1_0)^{-1}-\gamma A_{11}(0)$ is a positive definite matrix.
\end{asm}
Recall that $B^0:=(B^0_t)_{t\in[0,T]}$ is a $k$-dimensional $(\mathbb{F}^0,\mathbb{P}^0)$-standard Brownian motion independent of $W^0$. The SDE \eqref{LG_MRP} is well-posed due to the standard result for Lipschitz SDEs. The objective of this section is to find appropriate coefficients $(\alpha,\beta,\zeta,\eta)$ in \eqref{LG_MRP} with which $\theta$ satisfies \eqref{eqbm-theta}.
The following lemma shows that Assumption \ref{asm_kalman} (i) is consistent with Assumption \ref{asm1} (iv).

\begin{lem}~\\
    Under Assumption \ref{asm_kalman} (i), the Dol\'{e}ans-Dade exponential $\Bigl\{\mathcal{E}\Bigl(-\displaystyle\int_0^\cdot \theta_s^\top dW^0_s\Bigr)_t;t\in[0,T]\Bigr\}$ is a martingale.
\end{lem}
\noindent
\textbf{\textit{proof}}\\
We write:
\[
    \Lambda_t := \mathcal{E}\Bigl(-\displaystyle\int_0^\cdot \theta_s^\top dW^0_s\Bigr)_t,~~~t\in[0,T].
\]
By Bain \& Crisan \cite{bain_fundamentals_2009} [Lemma 3.9.], it suffices to show
\[
    \mathbb{E}\Bigl[\int_0^T |\theta_s|^2 ds\Bigr] <\infty,~~~\mathbb{E}\Bigl[\int_0^T  \Lambda_s|\theta_s|^2 ds\Bigr] <\infty.
\]
The first condition is obvious by the standard result for Lipschitz SDEs. The second condition can be shown similarly by following Bain \& Crisan \cite{bain_fundamentals_2009} [Exercise 3.11] and its solution in [Section 3.9].
$\square$

The observation is made according to the stock price process $(S_t)_{t\in[0,T]}$. By Lemma \ref{G=FW}, we can set
\begin{equation}
    \begin{split}
        \widetilde{W}^0_t = W^0_t + \int_0^t \theta_s ds
    \end{split}
\end{equation}
as an observation process. The dynamics of $\widehat\theta$ is given as follows.
\begin{lem}~\\
    \label{theta_hat_dynamics}
    Let Assumptions \ref{asm_kalman} (i) be in force. Then, the process $\widehat\theta$, defined by \eqref{theta_expected}, satisfies the following SDE:
    \begin{equation}
        \begin{split}
            \label{hat_theta}
            d\widehat\theta_t = (\alpha_t\widehat\theta_t + \beta_t)dt + (\zeta_t + \varrho_t) d\widehat{W}^0_t,~~~t\in[0,T],~~~\widehat\theta_0 = m,
        \end{split}
    \end{equation}
    where $\varrho \in\mathcal{C}^1([0,T];\mathbb{M}_{d_0})$ is a function which satisfies the following Riccati equation:
    \begin{equation}
        \begin{split}
            \label{ODE_for_vrho}
            \dot{\varrho }_t=\eta_t\eta_t^\top + \alpha_t \varrho_t + \varrho_t\alpha_t^\top - \zeta_t\varrho_t - \varrho_t\zeta_t - \varrho_t^2,~~~t\in[0,T],~~~\varrho_0 = v.
        \end{split}
    \end{equation}
\end{lem}
\noindent
\textbf{\textit{proof}}\\
See Liptser \& Shiryayev \cite{LiptserShiryayev} [Theorem 10.3] and set
\begin{equation}
    \begin{split}
    a_0=\beta,~~~a_1=\alpha,~~~a_2\equiv 0,~~~b_1=\zeta,~~~b_2=\eta,~~~A_0\equiv 0,~~~A_1\equiv I_{d_0},~~~A_2\equiv 0,~~~B_1\equiv I_{d_0},~~~B_2\equiv 0
    \end{split}
\end{equation}
therein. $\square$

In addition to Assumptions \ref{asm1} and \ref{asm2}, let Assumption \ref{asm_kalman} be in force. If $\widehat\theta$ satisfies
\begin{equation}
    \label{eqbm_hat_theta}
    \widehat\theta_t = -\gamma {\Sigma}_0^\top \Bigl(A_{00}(t)x^0_t + A_{10}(t)^\top \mu^1_t + B_0(t) \Bigr),~~~t\in[0,T],
\end{equation}
the processes $(p^{i,*})_{i\in\mathbb{N}}$ defined by
\[
    p^{i,*}_t := (\pi^{i,*}_t)^\top\sigma_t := Z^{i,0}_t + \frac{\widehat\theta_t^\top}{\gamma},~~~t\in[0,T],~~~i\in\mathbb{N},
\]
are optimal strategies and satisfy the asymptotic market clearing condition by Theorem \ref{EQG-verification}. Plugging \eqref{eqbm_hat_theta} into \eqref{hat_theta}, we get:
\begin{equation}
    \begin{split}
        \label{eq_3.3-1}
    d\widehat\theta_t 
    &=
    (\alpha_t \widehat\theta_t + \beta_t)dt + (\zeta_t + \varrho_t)d\widehat{W}^0_t\\
    &=
    \{-\gamma\alpha_t {\Sigma}_0^\top A_{00}(t)x_t^0 - \gamma\alpha_t {\Sigma}_0^\top (A_{10}(t)^\top \mu^1_t + B_0(t)) + \beta_t\}dt + (\zeta_t + \varrho_t)d\widehat{W}^0_t,~~~t\in[0,T],\\
    \widehat\theta_0 &= m = -\gamma {\Sigma}_0^\top \Bigl(A_{00}(0)x^0_0 + A_{10}(0)^\top \mathbb{E}[x^1_0] + B_0(0) \Bigr).
    \end{split}
\end{equation}
On the other hand, by applying Ito formula to \eqref{eqbm_hat_theta}, we have
\begin{equation}
    \begin{split}
        \label{eq_3.3-2}
    d\widehat\theta_t 
    &=
    -\gamma {\Sigma}_0^\top \{\dot{A}_{00}(t)x^0_tdt + A_{00}(t)dx^0_t +  \dot{A}_{10}(t)^\top \mu^1_t dt + A_{10}(t)^\top \dot{\mu}^1_t dt + \dot{B}_0(t) dt\}\\
    &=
    -\gamma {\Sigma}_0^\top\{(\dot{A}_{00}(t) - K_0 A_{00}(t))x^0_t + (K_0A_{00}(t)m_0 + \dot{A}_{10}(t)^\top \mu^1_t + A_{10}(t)^\top \dot{\mu}^1_t + \dot{B}_0(t) )\}dt\\
    &~~~~~   -\gamma {\Sigma}_0^\top A_{00}(t)\Sigma_0 d\widehat{W}_t^0,~~~t\in[0,T].
    \end{split}
\end{equation}
Comparing the coefficients of \eqref{eq_3.3-1} and \eqref{eq_3.3-2} with respect to the $x^0$-term and the constant term in the drift term as well as the diffusion term, we obtain
\begin{equation}
    \begin{split}
    &-\gamma\alpha_t {\Sigma}_0^\top A_{00}(t) = -\gamma {\Sigma}_0^\top(\dot{A}_{00}(t) - K_0 A_{00}(t)),\\
    &- \gamma\alpha_t {\Sigma}_0^\top (A_{10}(t)^\top \mu^1_t + B_0(t)) + \beta_t = -\gamma {\Sigma}_0^\top (K_0A_{00}(t)m_0 + \dot{A}_{10}(t)^\top \mu^1_t + A_{10}(t)^\top \dot{\mu}^1_t + \dot{B}_0(t) ),\\
    &\zeta_t + \varrho_t = -\gamma {\Sigma}_0^\top A_{00}(t)\Sigma_0
    \end{split}
\end{equation}
for each $t\in[0,T]$. Rearranging the terms, we get, for $t\in[0,T]$,
\begin{equation}
    \begin{split}
    &\alpha_t = \Sigma_0^\top \dot{A}_{00}(t)A^{-1}_{00}(t)(\Sigma_0^\top)^{-1} - K_0 I_{d_0}, \\
    &\beta_t = \gamma\alpha_t {\Sigma}_0^\top (A_{10}(t)^\top \mu^1_t + B_0(t)) -\gamma {\Sigma}_0^\top (K_0A_{00}(t)m_0 + \dot{A}_{10}(t)^\top \mu^1_t + A_{10}(t)^\top \dot{\mu}^1_t + \dot{B}_0(t) ),\\
    &\varrho_t = -\gamma {\Sigma}_0^\top A_{00}(t)\Sigma_0 - \zeta_t.
    \end{split}
\end{equation}

It is easy to see, for $t\in[0,T]$,
\begin{equation}
    \begin{split}
    &\dot\varrho_t = -\gamma {\Sigma}_0^\top \dot{A}_{00}(t)\Sigma_0 - \dot\zeta_t,\\
    &\alpha_t\varrho_t = -\gamma\Sigma^\top_0 (\dot A_{00}(t) - K_0 A_{00}(t))\Sigma_0 - \alpha_t\zeta_t,\\
    &\varrho_t\alpha_t^\top = -\gamma\Sigma^\top_0 (\dot A_{00}(t) - K_0 A_{00}(t))\Sigma_0 - \zeta_t\alpha_t^\top,\\
    &\zeta_t\varrho_t = -\gamma\zeta_t\Sigma^\top A_{00}\Sigma_0 - \zeta_t^2,\\
    &\varrho_t \zeta_t = -\gamma\Sigma^\top A_{00}\Sigma_0 \zeta_t - \zeta_t^2,\\
    &\varrho_t^2 = \gamma^2 \Sigma_0^\top A_{00}(t)\Sigma_0\Sigma_0^\top A_{00}(t)\Sigma_0 + \gamma\zeta_t\Sigma^\top A_{00}\Sigma_0 + \gamma\Sigma^\top A_{00}\Sigma_0 \zeta_t + \zeta_t^2.
    \end{split}
\end{equation}
By \eqref{ODE_for_vrho}, we have
\begin{equation}
    \begin{split}
        0
        &=
        \dot{\varrho }_t - \eta_t\eta_t^\top - \alpha_t \varrho_t - \varrho_t\alpha_t^\top + \zeta_t\varrho_t + \varrho_t\zeta_t + \varrho_t^2 \\
        &=
        -\gamma {\Sigma}_0^\top \dot{A}_{00}(t)\Sigma_0 - \dot\zeta_t - \eta_t\eta_t^\top \\
        &~~~+ \gamma\Sigma^\top_0 (\dot A_{00}(t) - K_0 A_{00}(t))\Sigma_0 + \alpha_t\zeta_t + \gamma\Sigma^\top_0 (\dot A_{00}(t) - K_0 A_{00}(t))\Sigma_0 + \zeta_t\alpha_t^\top \\
        &~~~-\gamma\zeta_t\Sigma^\top A_{00}\Sigma_0 - \zeta_t^2 -\gamma\Sigma^\top A_{00}\Sigma_0 \zeta_t - \zeta_t^2 \\
        &~~~+\gamma^2 \Sigma_0^\top A_{00}(t)\Sigma_0\Sigma_0^\top A_{00}(t)\Sigma_0 + \gamma\zeta_t\Sigma^\top A_{00}\Sigma_0 + \gamma\Sigma^\top A_{00}\Sigma_0 \zeta_t + \zeta_t^2\\
        &=
        - \eta_t\eta_t^\top - \dot\zeta_t + \alpha_t\zeta_t + \zeta_t\alpha_t^\top - \zeta_t^2\\
        &~~~+ \gamma\Sigma_0^\top (\dot A_{00}(t) - 2K_0 A_{00}(t) + \gamma  A_{00}(t)\Sigma_0\Sigma_0^\top A_{00}(t))\Sigma_0\\
        &=
        - \eta_t\eta_t^\top - \dot\zeta_t + \alpha_t\zeta_t + \zeta_t\alpha_t^\top - \zeta_t^2 - \gamma^2\Sigma_0^\top A_{10}(t)^\top \Sigma\Sigma^\top A_{10}(t)\Sigma_0.\\
    \end{split}
\end{equation}
Here, we used the ODE \eqref{Riccati eqn} in the last equality.\par
Above all, in order to make \eqref{hat_theta} consistent with \eqref{eqbm_hat_theta}, the initial condition $\widehat\theta_0$ and the coefficients $(\alpha_t,\beta_t,\rho_t)$ must be given by
\begin{equation}
    \begin{split}
    &\widehat\theta_0 = -\gamma {\Sigma}_0^\top \Bigl(A_{00}(0)x^0_0 + A_{10}(0)^\top \mathbb{E}[x^1_0] + B_0(0) \Bigr),\\
    &\alpha_t = \Sigma_0^\top \dot{A}_{00}(t)A^{-1}_{00}(t)(\Sigma_0^\top)^{-1} - K_0 I_{d_0},~~~t\in[0,T], \\
    &\beta_t = \gamma\alpha_t {\Sigma}_0^\top (A_{10}(t)^\top \mu^1_t + B_0(t)) -\gamma {\Sigma}_0^\top (K_0A_{00}(t)m_0 + \dot{A}_{10}(t)^\top \mu^1_t + A_{10}(t)^\top \dot{\mu}^1_t + \dot{B}_0(t) ),~~~t\in[0,T],\\
    &\varrho_t = -\gamma {\Sigma}_0^\top A_{00}(t)\Sigma_0 - \zeta_t,~~~t\in[0,T],
    \end{split}
\end{equation}
where $\zeta_t$ needs to satisfy the Riccati equation
\begin{equation}
    \begin{split}
        \label{coeff_ze_eta}
    &\dot\zeta_t = - \zeta_t^2  + \alpha_t\zeta_t + \zeta_t\alpha_t^\top  - \eta_t\eta_t^\top - \gamma^2\Sigma_0^\top A_{10}(t)^\top \Sigma\Sigma^\top A_{10}(t)\Sigma_0,~~~t\in[0,T],\\
    &\zeta_0 = -\gamma {\Sigma}_0^\top A_{00}(0)\Sigma_0 - v.
    \end{split}
\end{equation}
for $\eta\in\mathcal{C}([0,T];\mathbb{R}^{d_0\times k})$. These observations result in the following theorem.
\begin{thm} ~ \\
    Let Assumptions \ref{asm1}, \ref{asm2} and \ref{asm_kalman} be in force. Furthermore, assume that the mean $m(:=\mathbb{E}[\theta_0])$ and the coefficients $(\alpha,\beta,\zeta,\eta)\in\mathcal{C}([0,T];\mathbb{R}^{d_0\times d_0})\times \mathcal{C}([0,T];\mathbb{R}^{d_0}) \times \mathcal{C}^1([0,T];\mathbb{M}_{d_0})\times \mathcal{C}([0,T];\mathbb{R}^{d_0\times k})$ satisfy
    \begin{equation}
        \begin{split}
            \label{coeff_al_be}
        &m = -\gamma {\Sigma}_0^\top \Bigl(A_{00}(0)x^0_0 + A_{10}(0)^\top \mathbb{E}[x^1_0] + B_0(0) \Bigr),\\
        &\alpha_t = \Sigma_0^\top \dot{A}_{00}(t)A^{-1}_{00}(t)(\Sigma_0^\top)^{-1} - K_0 I_{d_0},~~~t\in[0,T], \\
        &\beta_t = \gamma\alpha_t {\Sigma}_0^\top (A_{10}(t)^\top \mu^1_t + B_0(t)) -\gamma {\Sigma}_0^\top (K_0A_{00}(t)m_0 + \dot{A}_{10}(t)^\top \mu^1_t + A_{10}(t)^\top \dot{\mu}^1_t + \dot{B}_0(t) ),~~~t\in[0,T],~~~~\\
        &\dot\zeta_t = - \zeta_t^2  + \alpha_t\zeta_t + \zeta_t\alpha_t^\top  - \eta_t\eta_t^\top - \gamma^2\Sigma_0^\top A_{10}(t)^\top \Sigma\Sigma^\top A_{10}(t)\Sigma_0,~~~t\in[0,T],\\
        &\zeta_0 = -\gamma {\Sigma}_0^\top A_{00}(0)\Sigma_0 - v.
        \end{split}
    \end{equation}
    and that such $\zeta$ is well-defined. Then, the asymptotic market clearing condition \eqref{MC-eqn} is satisfied as long as each agent adopts 
    \begin{equation}
       p^{i,*}_t := (\pi^{i,*}_t)^\top \sigma_t := Z^{i,0}_t + \frac{\widehat{\theta}_t^\top}{\gamma}, ~~~t\in[0,T], ~~~i\in\mathbb{N},
    \end{equation}
    as his/her optimal strategy. Here, $Z^{i,0}$ is given by \eqref{EQG solution YZ} and $\widehat{\theta}_t := \mathbb{E}[\theta_t|\mathcal{G}^0_t]$.
\end{thm}
\noindent
\textbf{\textit{proof}}\\
By Lemma \ref{theta_hat_dynamics}, $\widehat{\theta}$ follows \eqref{hat_theta}, where $\varrho \in\mathcal{C}^1([0,T];\mathbb{M}_{d_0})$ satisfies \eqref{ODE_for_vrho}. 
By the observation above, $\varrho_t = -\gamma {\Sigma}_0^\top A_{00}(t)\Sigma_0 - \zeta_t$ for $t\in[0,T]$ solves \eqref{ODE_for_vrho} and the local Lipschitz property implies that it is the unique solution. 
Then, the dynamics of $\widehat{\theta}$ reads
\[
    d\widehat\theta_t = (\alpha_t\widehat\theta_t + \beta_t)dt -\gamma {\Sigma}_0^\top A_{00}(t)\Sigma_0 d\widehat{W}^0_t,~~~t\in[0,T].
\]
Notice that the process $\widehat{\theta}$ satisfying above is unique due to the standard result for Lipschitz SDEs. By \eqref{eq_3.3-1} and \eqref{eq_3.3-2}, this clearly shows that $\widehat\theta$ is given by \eqref{eqbm_hat_theta}. 
The statement follows immediately from Theorem \ref{EQG-verification}. $\square$

\section{Numerical analysis}
In this section, we provide a numerical simulation to visualize the dynamics of our model. We consider an economy with $N=5000$ agents with time horizon $T=1$ and set $d_0 = d = k = 1$ for simplicity. Moreover, we set: \\
\begin{center}
\begin{tabular}{|c|c|c|c|c|c|c|c|c|c|c|c|c|c|c|c|} \hline
    $\gamma$ & $K_0$ & $K$ & $m_0$ & $m$ & $\Sigma_0$ & $\Sigma$ & $A^F_{00}$ & $A^F_{11}$ & $A^F_{10}$ & $B^F_0$ & $B^F_1$ & $C^F$  \\ \hline
    1.5 & 0.05 & 0.05 & $-0.5$ & $-0.5$ & 0.3 & 0.3 & 0.7 & 0.2 & 0.3 & $-1.3$ & $-0.7$ & 1.2 \\ \hline
\end{tabular}
\end{center}
Figure 1 presents the numerical solution of the ODEs \eqref{Riccati eqn}.
\begin{figure}[H] 
    \centering
    \includegraphics[width=17.5cm]{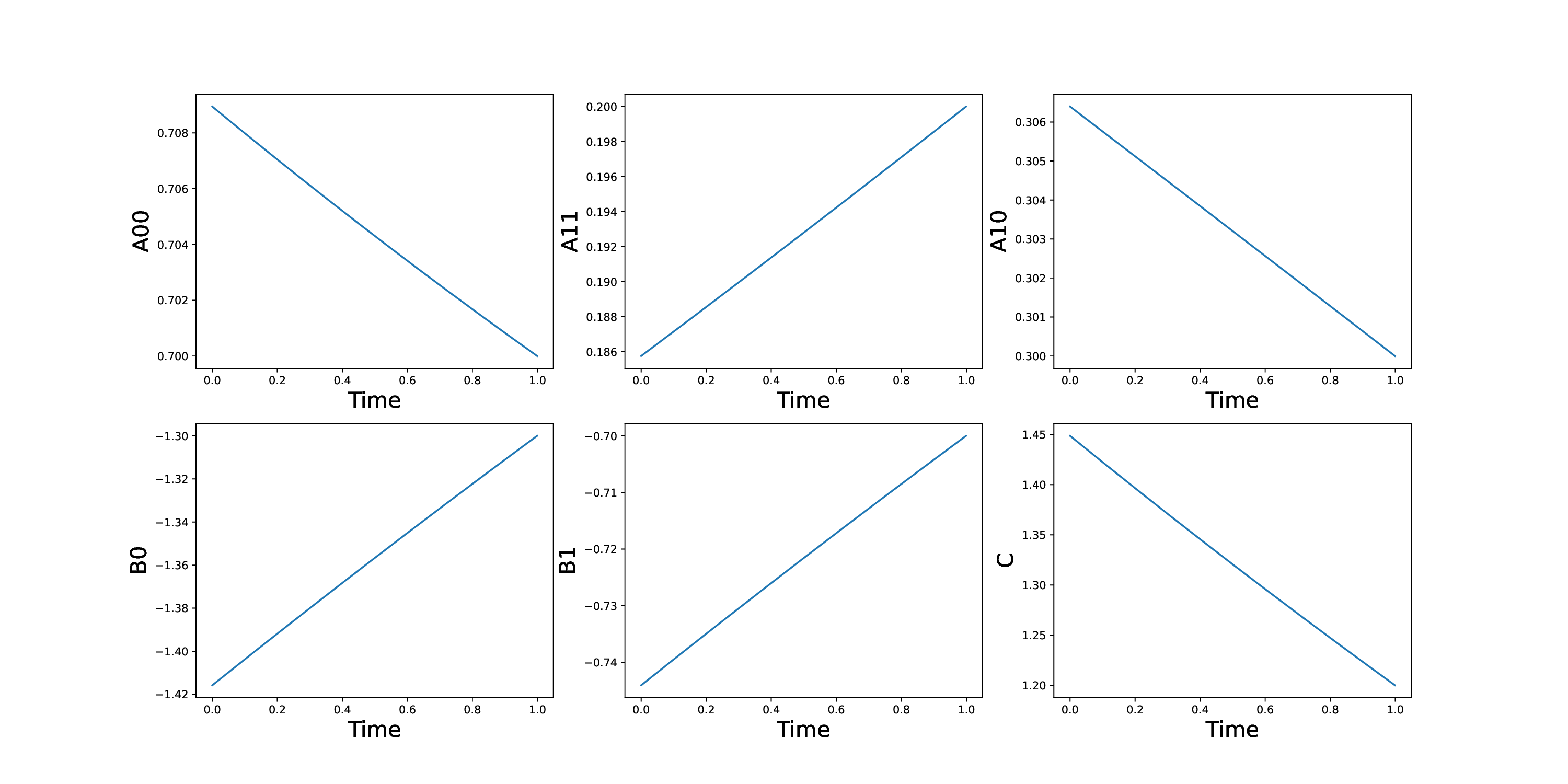}
    \caption{Solutions of Eq.\eqref{Riccati eqn}}
\end{figure}
We set $x^i_0 \sim N(-0.7,0.5)$, $\xi^i\sim N(2,0.3)$ for each $i=1,\ldots, 5000$, $v=0.1$, $x_0^0 = 0$, $\sigma_t\equiv 0.2$ and $\eta_t = (t-0.6)\mathbbm{1}_{[0.6,1]}(t)$. The sample paths for the risk premium process $\theta$ (blue solid line) and the estimated one $\widehat\theta$ (orange dashed line) are given in Figure 2. Moreover, Figure 3 draws $\displaystyle\frac{1}{N}\sum_{i=1}^N \pi^{i,*}_t$ and illustrates the asymptotic market clearing property.
\begin{figure}[H]
    \centering
    \begin{minipage}[b]{0.49\columnwidth}
        \centering
        \includegraphics[width=7.5cm]{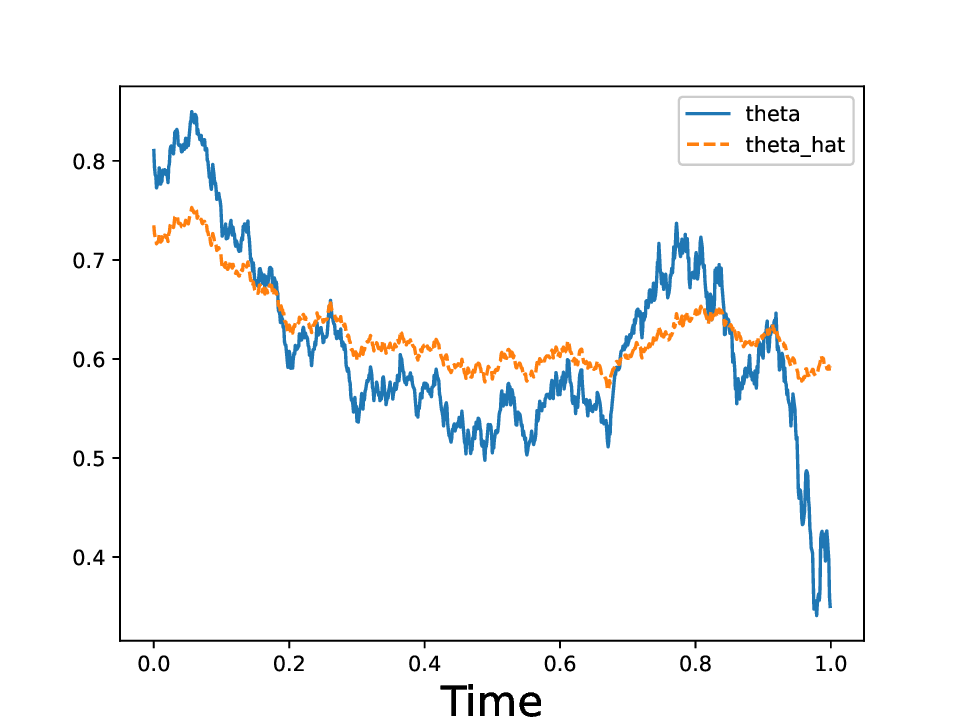}
        \caption{Market risk premium process}
    \end{minipage}
    \begin{minipage}[b]{0.49\columnwidth}
        \centering
        \includegraphics[width=7.5cm]{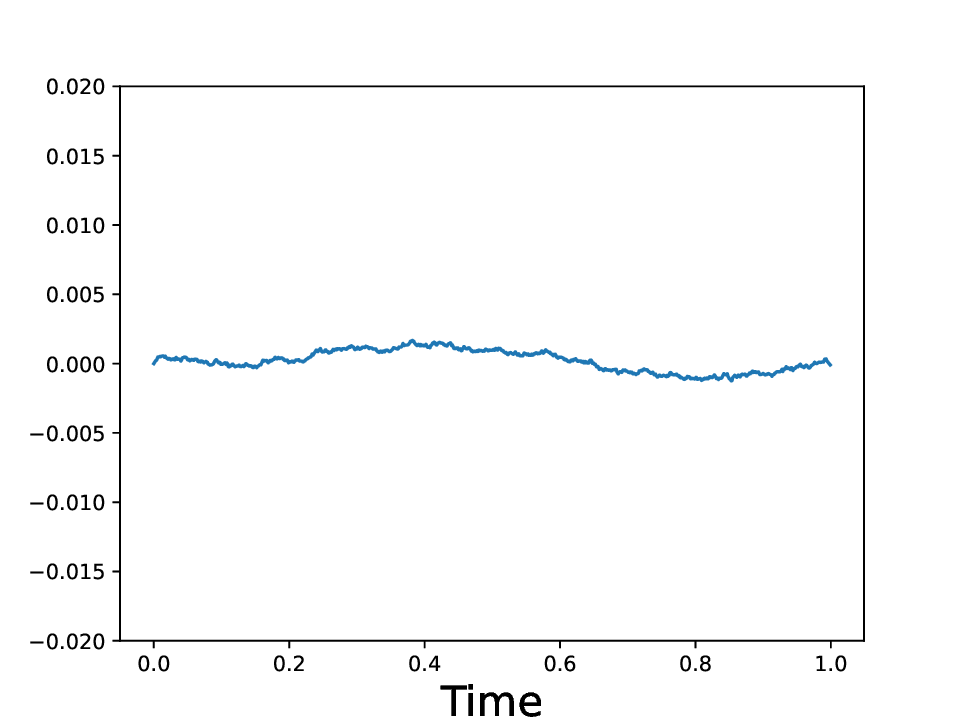}
        \caption{Asymptotic market clearing}
    \end{minipage}
\end{figure}

The distributions of agents' initial wealths $(\xi^i)_{i=1,\ldots,5000}$ and terminal wealths $(\mathcal{W}^{i,p^{i,*}}_T)_{i=1,\ldots,5000}$ are drawn in Figure 4 and terminal liabilities $(F^i)_{i=1,\ldots,5000}$ and terminal net assets $(\mathcal{W}^{i,p^{i,*}}_T-F^i)_{i=1,\ldots,5000}$ are drawn in Figure 5.

\begin{figure}[H] 
    \centering
    \includegraphics[width=17cm]{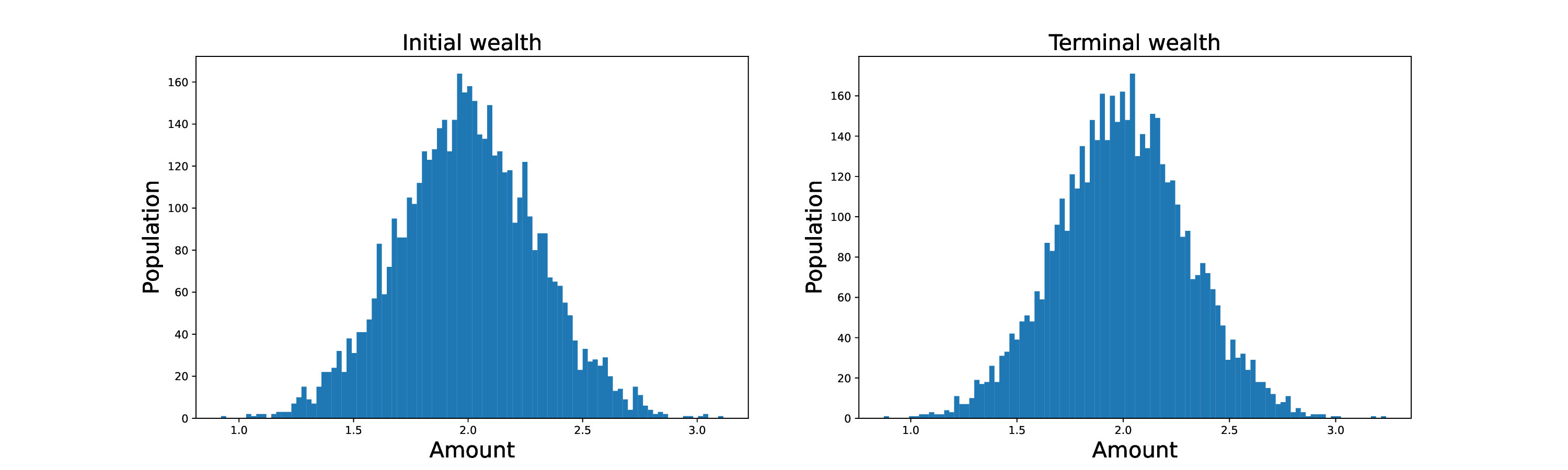}
    \caption{Initial and terminal wealth}
\end{figure}

\begin{figure}[H] 
    \centering
    \includegraphics[width=17cm]{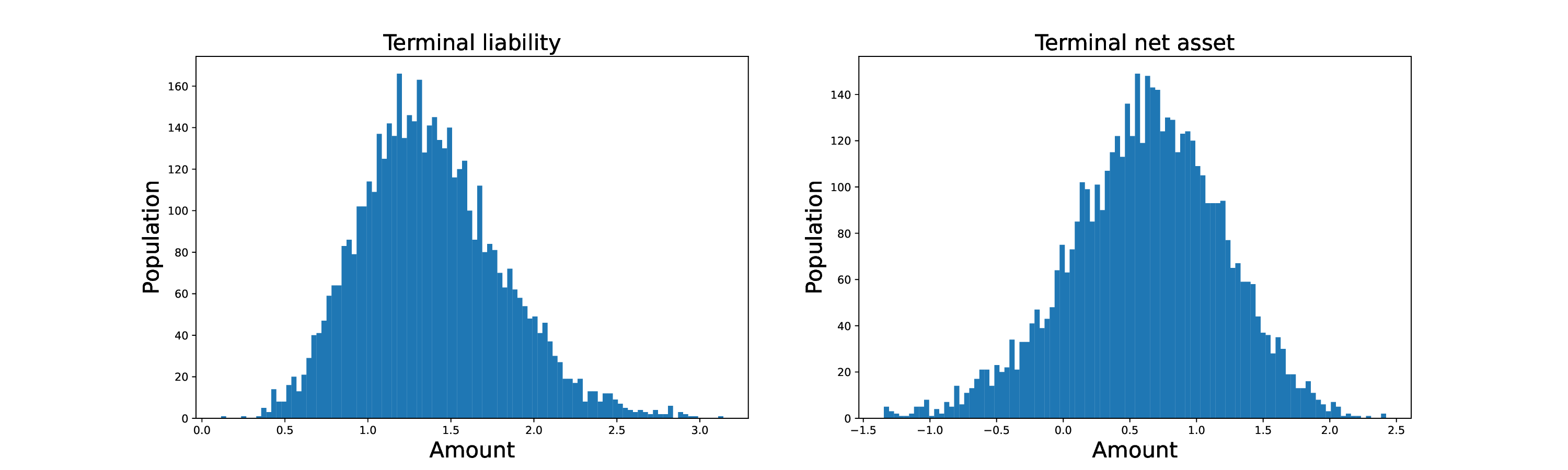}
    \caption{Terminal liability and net asset}
\end{figure}

\section{Conclusion and discussions}
In this paper, we studied the mean field equilibrium asset pricing model in a partially observable market. 
In Section 2, we formulated the utility maximization problem under partial observation and derived the condition for optimal strategies. 
In Section 3, within the exponential quadratic Gaussian framework, we associated the solution of the mean field BSDE with matrix ODEs and verified the asymptotic market clearing condition in the large population limit. 
We then constructed the risk premium process endogenously using the Kalman-Bucy filtering theory.
Section 4 presented a simple numerical example that visualizes a sample path of the risk premium process as well as distributions of agents' wealth.\par
As a direction for future research, we may possibly generalize the dynamics \eqref{LG_MRP} by, for example, adding jump process to formulate the possibility of default. This may lead us to consider the non-linear filtering for jump-diffusion processes.

\section*{Declarations}
\textbf{Funding}~~ The author is supported by Grant-in-Aid for JSPS Research Fellows, Grant Number JP23KJ0648.\\
\textbf{Conflicts of interests}~~ The author has no competing interests to declare that are relevant to this study.

\end{document}